\documentclass[aps,pre,reprint,superscriptaddress,amsmath,amssymb,floatfix,longbibliography]{revtex4-2}

\usepackage{graphicx}
\usepackage{dcolumn}
\usepackage{bm}
\usepackage[colorlinks=true,linkcolor=blue,citecolor=blue,urlcolor=blue]{hyperref}
\usepackage{physics}
\usepackage[version=4]{mhchem}
\usepackage{listings}
\usepackage{xcolor}

\newcommand{\eps}{\varepsilon}
\newcommand{\lB}{\ell_\text{B}}
\newcommand{\debye}{\lambda_\text{D}}
\newcommand{\lGC}{\ell_\text{GC}}
\newcommand{\lDu}{\ell_\text{Du}}
\newcommand{\rhoe}{\rho_\text{e}}
\newcommand{\Vs}{V_\text{s}}
\newcommand{\phis}{\phi_\text{s}}

\newcommand{\psim}{\psi_\text{m}}
\newcommand{\nm}{n_\text{m}}
\newcommand{\ns}{n_0} 

\newcommand{\ilm}{Universit{\'e} Claude Bernard Lyon 1, CNRS, Institut Lumière Matière, UMR5306, F69100 Villeurbanne, France}
\newcommand{\ill}{Institut Laue-Langevin, 71 Avenue des Martyrs, 38042 Grenoble, France}

\begin{document}


\title{Poisson-Boltzmann formulary: Third edition}

\author{Cecilia Herrero}
\email{cecilia.herreroguillen@gmail.com}
\affiliation{\ill}
\author{Laurent Joly}
\email{laurent.joly@univ-lyon1.fr}
\affiliation{\ilm}

\date{\today}

\begin{abstract}
The Poisson-Boltzmann (PB) equation provides a mean-field theory of electrolyte solutions at interfaces and in confinement, describing how ions reorganize close to charged surfaces to form the so-called electrical double layer (EDL), with numerous applications ranging from colloid science to biology. This formulary focuses on situations of interest for micro and nanofluidics, and gathers important formulas for the PB description of a $Z$:$Z$ electrolyte solution inside slit and cylindrical channels. Different approximated solutions (thin EDLs, no co-ion, Debye-H\"uckel, and  homogeneous/parabolic potential limits) and their range of validity are discussed, together with the full solution for the slit channel. Common boundary conditions are presented, the thermodynamics of the EDL is introduced, and an overview of the application of the PB framework to the description of electrokinetic effects is given. Finally, the limits of the PB framework are briefly discussed, and Python scripts to solve the PB equation numerically are provided. 
\end{abstract}

\maketitle

\tableofcontents

\section{Introduction}
\label{sec:intro}

\begin{figure}
    \centering
    \includegraphics[width=\linewidth]{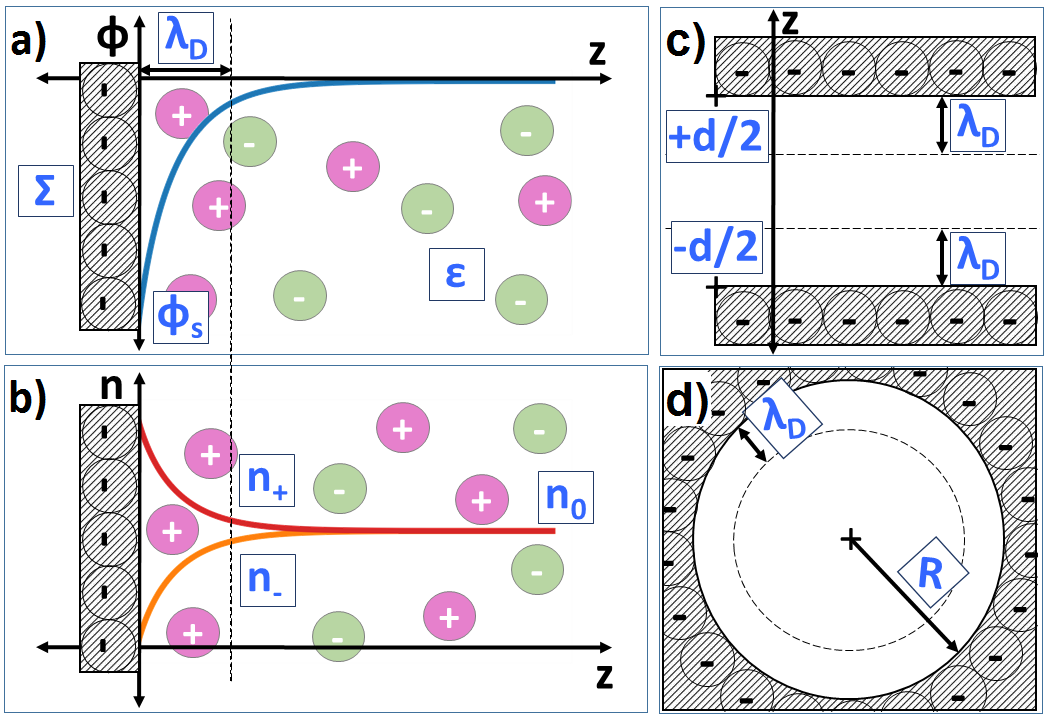}
    \caption{a-b) Local picture of the electrical double layer at a charged wall ($z$ is the distance from the wall, $\Sigma$ the surface charge density, $\eps$ the solvent dielectric permittivity, and $\debye$ the Debye length): a) Profile of reduced electric potential $\phi$, with $\phis$ the value at the wall; b) Profiles of cation $n_+$ and anion $n_-$ densities, with $n_0$ the bulk value. c) Schematics of the slit channel. d) Schematics of the cylindrical channel.}
    \label{fig:PB_figs}
\end{figure}

When an electrolyte solution meets a solid surface, several mechanisms can generate a surface charge, together with an opposite charge carried by ions in the liquid (typical mechanisms include dissociation of surface groups and specific adsorption of charged species) \cite{LyklemaBook,Hunter2001,IsraelachviliBook}.
In the vicinity of a charged wall, ions reorganize to form an oppositely charged region, which consists of two layers: 1. a layer of adsorbed ions, usually called the Stern or Helmoltz layer; 2. a diffuse layer of free ions, with counter-ions in excess and co-ions in defect, extending over the so-called Debye length denoted $\debye$, see Fig.~\ref{fig:PB_figs}(a-b). The overall interfacial distribution of charge, i.e. the charged surface, the adsorbed ions, and the diffuse layer, is generally referred to as the 
the electrical double layer (EDL), although this term is sometimes used to refer to the diffuse layer only. 
The EDL plays a key role in many aspects of soft condensed matter, as it controls the stability and dynamics of charged objects in solution  \cite{LyklemaBook,Hunter2001,IsraelachviliBook}. In particular, the EDL is at the origin of the so-called electrokinetic (EK) effects, where gradients and fluxes of different types (hydrodynamic, electrical, chemical, thermal) are coupled in the presence of charged interfaces \cite{Delgado2007}. Such EK effects are central to the very active fields of micro and nanofluidics \cite{Schoch2008,Bocquet2010,Hartkamp2018,Mouterde2019,Kavokine2021,Joly2021,Boon2022,Herrero2022,Robin2023,Sarma2024}. 

The equilibrium ion distribution and electric potential profile in the diffuse layer 
can be computed by combining the Poisson equation for electrostatics and the Boltzmann distribution of the ions into the so-called Poisson-Boltzmann (PB) equation, under certain assumptions \cite{Andelman1995,Markovich2016a}: 
\begin{itemize}
    \item the Poisson equation is written assuming that the solvent has a local, homogeneous and isotropic dielectric permittivity; 
    \item the equilibrium Boltzmann distribution of the ions is written assuming that the energy of the ions results only from their Coulomb interactions with the other ions and the wall, described at a mean-field level. 
\end{itemize} 
Educational presentations of the PB theory can be found in 
books 
\cite{LyklemaBook,Hunter2001,IsraelachviliBook,Blossey2023}, 
book chapters 
\cite{Andelman1995,Markovich2016a}, 
and articles 
\cite{Delgado2007}, 
discussing in particular applications to micro and nanofluidics 
\cite{Schoch2008,Bocquet2010,Hartkamp2018,Kavokine2021}.

In contrast, this formulary simply gathers important formulas for the description of the EDL with the PB equation, focusing on a $Z$:$Z$ electrolyte solution inside slit and cylindrical channels, see Fig.~\ref{fig:PB_figs}(c) and Fig.~\ref{fig:PB_figs}(d), respectively.
After introducing the notations (Sec.~\ref{sec:notations}) and important characteristic lengths (Sec.~\ref{sec:lengths}), the formulary describes the cases of a slit channel (Sec.~\ref{sec:slit}) and of a cylindrical channel (Sec.~\ref{sec:cyl}). For both geometries, different approximated solutions are discussed: the thin EDL limit, where $\debye$ is small as compared to the channel size; the no co-ion limit, where the EDLs overlap and the surface charge is large enough that co-ions are excluded; the Debye-H\"uckel limit at low surface charge, where the PB equation can be linearized; the homogeneous/parabolic potential limits, for large EDL overlap and low surface charge. The range of validity of these approximations is then discussed; for the slit channel only, the general solution is presented. 
Common boundary conditions are presented (Sec.~\ref{sec:BC}), the thermodynamics of the EDL is introduced (Sec.~\ref{sec:thermodynamics}), and an overview of the application of the PB theory to the description of EK effects is given (Sec.~\ref{sec:EK_effects}). Finally, the limits of the PB framework are briefly discussed (Sec.~\ref{sec:limits}). In the appendix, Python scripts to solve the PB equation numerically are provided.

\section{Abbreviations and notations}
\label{sec:notations}

\begin{itemize}
    \item PB: Poisson-Boltzmann
    \item EDL: electrical double layer
    \item EK: electrokinetic
\end{itemize}

\begin{itemize}
    \item solvent dielectric permittivity $\eps$
    \item inverse thermal energy $\beta = 1/(k_\text{B}T)$, with $k_\text{B}$ the Boltzmann constant and $T$ the temperature
    \item absolute ionic charge $q = Ze$, with $e$ the elementary charge and $Z$ the ion valence
    \item ion number densities $n_\pm$
    \item salt concentration (number density) $n_0 = n_+ = n_-$ in the bulk or in the reservoirs, where the electrostatic potential vanishes
    \item free charge density $\rhoe$ (carried by the ions) 
    \item surface charge density $\Sigma$
    \item electrostatic potential $V$
    \item reduced potential $\phi = \beta q V$
    \item their value at the wall $\Vs$ and $\phis$
    \item $\gamma = \tanh \left( \phis / 4 \right)$
    \item auxiliary reduced potential $\psi = -\text{sgn}(\Sigma) \times \phi$
    \item Bjerrum length $\lB$
    \item Debye length $\debye$
    \item Gouy-Chapman length $\lGC$
    \item dimensionless surface charge $\chi = \debye / \lGC$
    \item Dukhin length $\lDu$
    \item electrostatic energy $\mathcal{E}$
\end{itemize}

\section{Characteristic lengths}
\label{sec:lengths}

To describe any $Z$:$Z$ electrolyte with the same formulas, the characteristic lengths in this formulary are defined based on the ionic charge $q=Ze$ rather than on the elementary charge $e$.

\paragraph*{Solvent permittivity $\eps$: Bjerrum length $\lB$ --}

In this formulary, the Bjerrum length is defined as
the distance at which the electrostatic interaction energy between two ions (of charge $q=Ze$) is equal to the thermal energy $k_\text{B}T$.
\begin{equation}\label{eq:lB}
\lB = \frac{\beta q^2}{4\pi \eps}
\quad \Leftrightarrow \quad  
\eps = \frac{\beta q^2}{4\pi \lB}
\end{equation}
For a monovalent salt in water at room temperature, $\lB \sim 0.7$\,nm.

\bigskip
\paragraph*{Bulk salt concentration $n_0$: Debye length $\debye$ --}

The Debye length is the range of the exponential screening of the electric field in an electrolyte. 
\begin{equation}\label{eq:debye}
\debye = \frac{1}{\sqrt{8\pi \lB n_0}}
= \sqrt{\frac{\eps}{2\beta q^2 n_0}} 
\quad \Leftrightarrow \quad 
n_0 = \frac{1}{8\pi \lB \debye^2}
\end{equation}
For a monovalent salt in water at room temperature, $\debye \sim 0.3\,\text{nm} / \sqrt{c_0 \text{(mol/L)}}$, where $c_0 = n_0 / N_\text{A}$ is the molar salt concentration, with $N_\text{A}$ the Avogadro constant.

\bigskip
\paragraph*{Surface charge density $\Sigma$: Gouy-Chapman length $\lGC$ --}

In this formulary, the Gouy-Chapman length is defined as
the distance at which the electrostatic interaction energy between an ion (of charge $q=Ze$) and a charged surface is comparable to the thermal energy $k_\text{B}T$.
\begin{equation}\label{eq:lGC}
\lGC = \frac{q}{2\pi \lB |\Sigma|}
= \frac{2\eps}{\beta q |\Sigma|} 
\quad \Leftrightarrow \quad 
\Sigma = \frac{q\ \text{sgn}(\Sigma)}{2\pi \lB \lGC}
\end{equation}
For monovalent ions in water at room temperature, $\lGC \sim 36\,\text{nm} / |\Sigma| \mathrm{(mC/m^2)}$.

A useful quantity is the dimensionless surface charge 
$\chi = \debye/\lGC$, which is proportional to the absolute value of the surface charge, and inversely proportional to the square root of the bulk salt concentration: 
\begin{equation*}
\chi = \frac{\debye}{\lGC} = 2\pi \lB \debye \frac{|\Sigma|}{q} 
= \sqrt{\frac{\beta \Sigma^2}{8 \eps \ns}}
\end{equation*}

\bigskip
\paragraph*{Surface ions vs bulk ions: Dukhin length $\lDu$ --}

The Dukhin length is the channel scale at which the number of salt ions in bulk compares to the number of counter-ions at the surfaces \cite{Bocquet2010}. 
\begin{equation}\label{eq:lDu}
    \lDu = \frac{\abs{\Sigma}/q}{n_0} 
\end{equation}

\section{Slit channel}
\label{sec:slit}

This section gathers the equations describing a $Z$:$Z$ electrolyte solution confined between two planar parallel walls at a distance $d$, see Fig.~\ref{fig:PB_figs}(c). 

\subsection{Thin EDLs: one wall with salt}
\label{sec:onewall_salt}

When the distance $d$ between the surfaces is much larger than the thickness of the EDLs, one can solve the PB equation for a single charged wall, and superimpose the potentials of the two walls to obtain the full potential in the channel. The limits of validity of this approximation will be quantified in section~\ref{sec:slit_r1r2}. 

Accordingly, this section gathers formulas for the description of a $Z$:$Z$ electrolyte solution in contact with a single planar wall, where $z \geq 0$ is the distance from the wall, see Fig.~\ref{fig:PB_figs}(a) and (b).

\subsubsection{PB equation: derivation}

\paragraph*{Poisson equation:} 
\begin{equation}\label{eq:Poisson_onewall}
    -\eps \dfrac{\mathrm{d}^2 V}{\mathrm{d} z^2} = \rhoe(z) = q \{n_+(z) - n_-(z)\}
\end{equation}

\paragraph*{Boltzmann distribution:} 
\begin{equation*}
    n_\pm (z) = n_0\,\exp{\mp \beta q V(z)}
\end{equation*}

\paragraph*{Poisson-Boltzmann equation} 
\begin{equation}\label{eq:PB_onewall}
\frac{\mathrm{d}^2 \phi}{\mathrm{d} z^2} = \frac{1}{\debye^2} \sinh\{\phi(z)\} 
\end{equation}

\subsubsection{PB equation: solution} 

Multiplying both sides of Eq.~\eqref{eq:PB_onewall} by $\mathrm{d}\phi/\mathrm{d}z$ and integrating between $z=\infty$ and $z$; imposing that $\mathrm{d}\phi/\mathrm{d}z|_{z=\infty} = \phi(\infty) = 0$: 
\begin{equation}\label{eq:coshphi_onewall}
\frac{1}{2} \left(\frac{\mathrm{d} \phi}{\mathrm{d} z} \right)^2 = \frac{1}{\debye^2} \left\{ \cosh(\phi) - 1 \right\} 
\end{equation}
Noting that $\cosh(\phi)-1=2\sinh^2(\phi/2)$, and that $\phi$ and $\mathrm{d}\phi/\mathrm{d}z$ are of opposite sign: 
\begin{equation}\label{eq:d_z_phi_onewall}
\frac{\mathrm{d} \phi}{\mathrm{d} z}  = -\frac{2}{\debye} \sinh\left(\frac{\phi}{2}\right)  
\end{equation}
Noting that $\sinh(\phi/2) = 2 \sinh(\phi/4) \cosh(\phi/4) = 2 \tanh(\phi/4) \cosh^2(\phi/4)$, and integrating like a physicist: 
\begin{multline*}
\int_{\phis}^\phi \frac{\mathrm{d}(\phi/4)}{\tanh(\phi/4) \cosh^2(\phi/4)} 
= - \int_0^z \frac{\mathrm{d}z}{\debye} \\ 
\quad \Rightarrow \quad 
\ln\left| \tanh\left(\frac{\phi}{4}\right) \right| - \ln\left| \tanh\left(\frac{\phis}{4}\right) \right|
= -\frac{z}{\debye}  
\end{multline*}
Noting that $\phi$ and $\phis$ have the same sign: 
\begin{equation*}
\tanh\left(\frac{\phi}{4}\right) = \tanh\left(\frac{\phis}{4}\right) e^{-\frac{z}{\debye}}
\end{equation*}

Noting that $\text{atanh}(x)=\frac{1}{2}\ln\left(\frac{1+x}{1-x}\right)$, and defining 
\begin{equation}
\gamma = \tanh\left(\frac{\phis}{4}\right)
\end{equation}
one finally gets: 
\begin{equation}\label{eq:phi_PB_onewall}
\phi(z) = 4 \,\text{atanh} \left\{ \gamma e^{-\frac{z}{\debye}} \right\} 
= 2\ln\left(\frac{1+\gamma e^{-\frac{z}{\debye}}}{1-\gamma e^{-\frac{z}{\debye}}}\right)
\end{equation}

\subsubsection{Surface charge and surface potential}

Electrostatics at the interface: 
\begin{multline}\label{eq:interface_ES_onewall}
- \left.\frac{\mathrm{d}V}{\mathrm{d}z}\right|_{z=0} = \frac{\Sigma}{\eps} \\
\quad \Rightarrow \quad 
\left.\frac{\mathrm{d}\phi}{\mathrm{d}z}\right|_{z=0} = - 4 \pi \lB \,\frac{\Sigma}{q} = - \frac{2 \,\text{sgn}(\Sigma)}{\lGC}
\end{multline}
Note that Eq.~\eqref{eq:interface_ES_onewall} also follows from the global electroneutrality of the system:
\begin{equation*}
\int_0^\infty \rhoe \dd z = -\int_0^\infty \eps \frac{\dd^2 V}{\dd z^2} \dd z = \eps \left.\frac{\mathrm{d}V}{\mathrm{d}z}\right|_{z=0} = -\Sigma
\end{equation*} 

Using Eq.~\eqref{eq:d_z_phi_onewall}, one obtains \emph{Grahame equation} relating $\phis$ and $\Sigma$: 
\begin{equation}\label{eq:grahame_onewall}
\sinh\left(\frac{\phis}{2}\right) 
= 2\pi \lB \debye \,\frac{\Sigma}{q} 
= \text{sgn}(\Sigma) \chi
\end{equation} 
so that $\phis$ can be expressed as: 
\begin{multline}\label{eq:grahame_phi0_onewall}
\phis
= 2 \,\text{sgn}(\Sigma) \,\text{asinh}\left( \chi \right)  \\ 
= 2 \,\text{sgn}(\Sigma) \ln \left\{ \chi + \sqrt{1+ \chi^2 } \right\}
\end{multline} 

One can also express $\gamma = \tanh(\phis/4)$ as a function of $\Sigma$, or equivalently as a function of $\chi$: 
\begin{equation}\label{eq:gamma_onewall}
\gamma = \frac{\text{sgn}(\Sigma)}{\chi} \left\{ -1 + \sqrt{1+\chi^2} \right\} 
\end{equation}

\subsubsection{Properties of the EDL and related integrals}
\label{sec:onewall_salt_integrals}

\paragraph{Electrostatic energy $\mathcal{E}$}

The electrostatic energy per unit area writes:
\begin{multline}\label{eq:E_onewall}
\mathcal{E} 
= \frac{\eps}{2} \int_0^\infty \left( \frac{\mathrm{d}V}{\mathrm{d}z} \right)^2 \,\mathrm{d}z 
= \frac{1}{8\pi \lB \beta} \int_0^\infty \left( \frac{\mathrm{d}\phi}{\mathrm{d}z} \right)^2 \,\mathrm{d}z \\
= \frac{1}{2\pi \lB \debye \beta} \left\{ -1 + \sqrt{1+\chi^2} \right\} 
= \frac{|\gamma|}{2\pi \lB \lGC \beta} 
\end{multline}

\paragraph{Ionic densities} 

\begin{itemize}
\item[$\circ$] \emph{
Surface excess / Adsorption $\Gamma$}

One can compute the excess number of ions per unit surface, commonly denoted $\Gamma$ (not to be mixed with the plasma parameter defined in section~\ref{sec:limits}):

Using Eq.~\eqref{eq:coshphi_onewall}, 
\begin{equation*}
n_+ + n_- - 2 n_0 
= 2n_0 \left\{ \cosh(\phi) -1 \right\} = 
\frac{1}{8\pi\lB} \left(\frac{\mathrm{d} \phi}{\mathrm{d} z} \right)^2 
\end{equation*} 
one obtains:
\begin{equation}\label{eq:dens_excess_onewall}
\Gamma = \int_0^\infty \left( n_+ + n_- - 2n_0\right) \,\mathrm{d}z 
= \beta \mathcal{E}
\end{equation}

\item[$\circ$] \emph{Density difference}

Electroneutrality imposes that: 
\begin{equation*}
\int_0^\infty \rhoe(z) \,\mathrm{d}z 
= - \Sigma 
\end{equation*}

Remembering that $\rhoe = q (n_+ - n_-)$, one gets:  
\begin{equation}\label{eq:dens_diff_onewall}
\int_0^\infty (n_+ - n_-) \,\mathrm{d}z 
= - \frac{\Sigma}{q} = - \frac{\text{sgn}(\Sigma)}{2\pi \lB \lGC} 
\end{equation} 
\end{itemize}

\paragraph{Differential capacitance} 

\begin{equation}
    C = \frac{\dd \Sigma}{\dd \Vs} = \frac{\eps}{\debye} \cosh\left( \frac{\phis}{2} \right) = \frac{\eps}{\debye} \sqrt{1+\chi^2}
\end{equation}

\paragraph{Surface tension} 

As detailed in Sec.~\ref{sec:thermodynamics}, the surface tension of the EDL, $\gamma_\text{EDL}$, is: 
\begin{equation}
    \gamma_\text{EDL} = -2\mathcal{E}
\end{equation}

\paragraph{More integrals}

The following integrals are useful for the calculation of EK response coefficients, see section~\ref{sec:EK_effects}. 

\begin{multline}\label{eq:calF_onewall}
\mathcal{F} = \int_0^\infty \left( \phis - \phi \right) \left( \frac{\mathrm{d}\phi}{\mathrm{d}z} \right)^2 \,\mathrm{d}z \\
=  \frac{8 \,\text{sgn}(\Sigma)}{\debye} \left\{ \chi - \text{asinh}(\chi) \right\}
\end{multline} 

\begin{equation}\label{eq:calG_onewall}
\mathcal{G} = \int_0^\infty z \left( \frac{\mathrm{d}\phi}{\mathrm{d}z} \right)^2 \,\mathrm{d}z 
= -4 \ln \left( 1 - \gamma^2 \right) 
\end{equation}

\subsubsection{Thickness of the EDL}

The thickness of the EDL is often identified with the Debye length $\debye$. However, at high $|\Sigma|$, the charge of the EDL is concentrated in a region much thinner than $\debye$. The characteristic thickness of the charged region can be defined as: 
\begin{equation*}
\lambda = \frac{\int_0^\infty z \,\rhoe(z) \,\mathrm{d}z}{\int_0^\infty \rhoe(z) \,\mathrm{d}z}
= \frac{\eps \Vs}{\Sigma} 
= \frac{\Vs}{-\left.\frac{\mathrm{d}V}{\mathrm{d}z}\right|_{z=0}} 
= \frac{\phis}{-\left.\frac{\mathrm{d}\phi}{\mathrm{d}z}\right|_{z=0}} 
\end{equation*}

Using Eq.~\eqref{eq:interface_ES_onewall} and Eq.~\eqref{eq:grahame_phi0_onewall}, one gets: 
\begin{equation}\label{eq:lambda_onewall}
\lambda = \debye \frac{\text{asinh}\left( \chi \right)}{\chi} = \lGC\, \text{asinh}\left( \chi \right)
\end{equation}

The characteristic scale for the variation of the electric field can also be written, following \citet{Weisbuch1983}: 
\begin{equation}\label{eq:lambdaprime_onewall}
\lambda' 
= \frac{\left. -\frac{\mathrm{d}\phi}{\mathrm{d}z}\right|_{z=0}}{\left.\frac{\mathrm{d}^2\phi}{\mathrm{d}z^2}\right|_{z=0}} 
= \frac{\debye}{\sqrt{1+ \chi^2 }}
\end{equation}

The limits at low and high surface charge of these characteristic scales are reported in section~\ref{sec:onewall_highlowSigma} (Table~\ref{tab:onewall_salt_limits}).

\subsubsection{Low and high surface charge limits}

\paragraph{Critical surface charge}

When the surface charge $|\Sigma|$ is low enough that the reduced potential $|\phi|$ is much lower than 1 everywhere, and therefore when $|\phis| = \max(|\phi|) \ll 1$, the PB equation can be linearized; this is the Debye-H\"uckel (DH) regime. 

Using Grahame equation, Eq.~\eqref{eq:grahame_onewall}, it appears that this regime is found when $\chi = \debye/\lGC \ll 1$.
The related critical surface charge for which $\debye = \lGC$ writes: 
\begin{equation}
|\Sigma|_\text{c} = \frac{q}{2\pi \lB \debye} = \sqrt{\frac{8\eps n_0}{\beta}}
\end{equation} 

For instance, for a monovalent salt in water at room temperature: 
\begin{equation}
|\Sigma|_\text{c} 
\sim \frac{36\,\text{mC/m}^2}{\debye \text{(nm)}} \sim 120\,\text{mC/m}^2 \sqrt{n_0 \text{(mol/L)}}
\end{equation} 

In practice, the linearized equation only provides a fair description for $|\Sigma| \lesssim 0.2 |\Sigma|_\text{c} = 24\,\text{mC/m}^2 \sqrt{n_0 \text{(mol/L)}}$. Thus, except at very high salt concentration, the DH regime is only found for very low surface charge. 

\bigskip
\paragraph{Debye-H\"uckel regime}

When $\chi = \debye/\lGC \ll 1$, the PB equation can be linearized and its solutions simplified: 
\begin{itemize}
    \item the PB equation becomes: $\dfrac{\mathrm{d}^2 \phi}{\mathrm{d} z^2} = \dfrac{1}{\debye^2} \phi$
    \item the potential becomes: $\phi(z) = \phis \,\text{e}^{-z/\debye}$
    \item $\gamma$ becomes $\phis/4$
    \item the Grahame relation becomes: $\Vs/\debye = \Sigma/\eps$ 
\end{itemize}

\bigskip
\paragraph{Some limits at low and high surface charge}
\label{sec:onewall_highlowSigma}

The limits of $\phis$, $\gamma$, $\mathcal{E}$, $\lambda$, $\lambda'$, $\mathcal{F}$ and $\mathcal{G}$ at low and high surface charge are reported in Table~\ref{tab:onewall_salt_limits}.

\begin{table}
\begin{center}
    \begin{tabular}{c|c|c}
        Quantity & $\chi = \debye/\lGC \ll 1$ & $\chi = \debye/\lGC \gg 1$ \\
        \hline
        $\phis$ & $\text{sgn}(\Sigma) 2\chi = \beta q \dfrac{\Sigma \debye}{\eps}$ 
        & $2\,\text{sgn}(\Sigma) \ln\left( 2\chi \right)$ \\[7pt]
        \hline
        $\gamma$ & $\text{sgn}(\Sigma) \dfrac{\chi}{2} = \beta q \dfrac{\Sigma \debye}{4 \eps}$ 
        & $\text{sgn}(\Sigma)$ \\[7pt]
        \hline
        $\mathcal{E}$ & $\dfrac{\chi^2}{4\pi \beta \lB \debye} = \dfrac{\Sigma^2 \debye}{4\eps}$ & $\dfrac{\chi}{2\pi \beta \lB \debye} = \dfrac{|\Sigma|}{\beta q}$ \\[7pt]
        \hline
        $C$ & $\dfrac{\eps}{\debye}$ & $\dfrac{\eps}{\lGC}$ \\[7pt]
        \hline
        $\lambda$ & $\debye$ & $\lGC \ln\left( 2\chi \right)$ \\[7pt]
        \hline
        $\lambda'$ & $\debye$ & $\lGC$ \\[7pt]
        \hline
        $\mathcal{F}$ & $\dfrac{4\, \text{sgn}(\Sigma) \chi^3}{3\debye}$  & $\dfrac{8 \,\text{sgn}(\Sigma) \chi}{\debye}$ \\[7pt]
        \hline
        $\mathcal{G}$ & $\chi^2$ & $4 \ln \left(\dfrac{\chi}{2}\right)$
    \end{tabular}
\end{center}
    \caption{One wall with salt: limits of $\phis$, $\gamma$, $\mathcal{E}$, $C$, $\lambda$, $\lambda'$, $\mathcal{F}$ and $\mathcal{G}$ for $\chi = \debye/\lGC \ll 1$ (low surface charge, DH regime) and for $\chi = \debye/\lGC \gg 1$ (high surface charge regime).}
    \label{tab:onewall_salt_limits}
\end{table}

\subsection{Thick EDLs, high surface charge: two walls, no co-ion}
\label{sec:twowalls_nosalt}

When the EDLs overlap and for large enough surface charges (these conditions will be quantified in section~\ref{sec:slit_r1r2}), co-ions are excluded from the channel and the PB equation can be solved with counter-ions only. 

\subsubsection{PB equation: derivation}

Let's consider counter-ions with density $n(z)$ confined between two parallel walls located at $z=-d/2$ and $z=d/2$, baring the same surface charge density $\Sigma$. 

In order to make the equations describing a negative surface charge (with positive counter-ions) or a positive surface charge (with negative counter-ions) identical, one can define an auxiliary reduced potential $\psi$, which will always be negative: 
\begin{equation}
    \psi = -\text{sgn}(\Sigma) \times \phi = -\text{sgn}(\Sigma) \times \beta q V .
\end{equation}

The potential in the middle of the channel is arbitrarily fixed to zero: $\psim = \psi(z=0) = 0$ (this condition will be relaxed in section~\ref{sec:slit_r1r2_nocoion}). Denoting $\nm$ the counter-ion density at $z=0$, and introducing a new characteristic length $K^{-1}$,
\begin{equation}
    K^{-1} = \frac{1}{\sqrt{2 \pi \lB \nm}} 
    \quad \Leftrightarrow \quad 
    \nm = \frac{K^2}{2\pi \lB}, 
\end{equation}
one can derive the PB equation. 

\bigskip
\paragraph*{Poisson equation:}
\begin{equation*}
    q n(z) = -\dfrac{\eps}{\beta q} \dfrac{\mathrm{d}^2 \psi}{\mathrm{d} z^2}
\end{equation*}

\paragraph*{Boltzmann distribution:}
\begin{equation*}
    n(z) = \nm\,\mathrm{e}^{-\psi(z)}
\end{equation*}

\paragraph*{Poisson-Boltzmann equation:}
\begin{equation}
    \dfrac{\mathrm{d}^2 \psi}{\mathrm{d} z^2} = -2 K^2 \,\mathrm{e}^{-\psi(z)}
\end{equation}

\subsubsection{PB equation: solution} 

\begin{equation}\label{eq:psi_twowalls_nosalt}
    \psi(z) = \ln\left\{\cos^2(K z)\right\}
\end{equation}
\begin{equation}\label{eq:dz_psi}
    \dfrac{\mathrm{d}\psi}{\mathrm{d}z} = -2 K \tan(Kz)
\end{equation}
\begin{equation}\label{eq:n_twowalls_nosalt}
    n(z) = \dfrac{\nm}{\cos^2(K z)}  = \dfrac{K^2}{2 \pi \lB \cos^2(K z)}
\end{equation}

\subsubsection{Surface charge and surface potential}

To fully determine the potential and ion density profiles, one needs to express $K$ as a function of the surface charge. 

Electrostatics at the interface (or equivalently, global electroneutrality): 
\begin{equation}
\left.\frac{\mathrm{d}V}{\mathrm{d}z}\right|_{z=d/2} = \frac{\Sigma}{\eps} 
\quad \Rightarrow \quad 
\left.\frac{\mathrm{d}\psi}{\mathrm{d}z}\right|_{z=d/2} = - \frac{2}{\lGC}
\end{equation}

Using Eq.~\eqref{eq:dz_psi}, one obtains: 
\begin{equation}\label{eq:Kd_vs_doverlGC}
Kd \,\tan\left(\frac{Kd}{2}\right) = \frac{d}{\lGC}.
\end{equation} 
Equation~\eqref{eq:Kd_vs_doverlGC} only provides an implicit expression for $K$, but
a very accurate explicit approximation (error below $0.07$\,\%) can be written: 
\begin{equation}
    \frac{Kd}{2} \approx \frac{\pi}{2} \sqrt{\frac{x (x+5)}{x^2+9x+5\pi^2/2}}, 
\end{equation}
with $x=d/\lGC$. 

Simpler approximate expressions can also be derived in the low and high surface charge limits, see Sec.~\ref{subsec:lowandhigsigmaSLIT}.

\subsubsection{A few integrals}

\paragraph{Electrostatic energy $\mathcal{E}$}

\begin{multline}
\mathcal{E} 
= \frac{\eps}{2} \int_{-d/2}^{d/2} \left( \frac{\mathrm{d}V}{\mathrm{d}z} \right)^2 \,\mathrm{d}z \\
= \frac{2}{\pi d \lB \beta} \frac{Kd}{2} \left[ \tan\left( \frac{Kd}{2} \right) - \frac{Kd}{2} \right] 
\end{multline} 
An approximate expression as a function of $x=d/\lGC$ (error below $1$\,\%) can be written: 
\begin{equation}
    \mathcal{E} \approx \frac{1}{\pi d \lB \beta} \times  \frac{x^2}{6-\sqrt{x}/5+x}. 
\end{equation}

\paragraph{Ionic density} 

\begin{equation}
\int_{-d/2}^{d/2} n(z) \,\mathrm{d}z 
= \frac{2|\Sigma|}{q} = \frac{1}{\pi \lB \lGC} 
\end{equation}

\subsubsection{Low and high surface charge limits}\label{subsec:lowandhigsigmaSLIT}

\paragraph{Low surface charge}
When $\Sigma$ is small enough that $d/\lGC \ll 1$, Eq.~\eqref{eq:Kd_vs_doverlGC} simplifies:
\begin{equation}\label{eq:Kd_vs_doverlGC_IG}
Kd \approx \sqrt{2 d/\lGC},
\end{equation} 
which is accurate within 1\,\% up to $d/\lGC\approx 0.12$. 

Note that when $d/\lGC \ll 1$, $Kd \ll 1$, so that $\psi \approx 0$ and the ion density is approximately homogeneous in the channel: 
\begin{equation}
    n \approx \dfrac{K^2}{2 \pi \lB} = \dfrac{1}{\pi d \lB \lGC} . 
\end{equation} 
This is commonly called the \emph{ideal gas} regime. In practice, the ion density varies by less than 10\,\% over the channel thickness as long as $d/\lGC < 0.2$. 

\bigskip
\paragraph{High surface charge}

When $\Sigma$ is large enough that $d/\lGC \gg 1$, Eq.~\eqref{eq:Kd_vs_doverlGC} simplifies:
\begin{equation}\label{eq:Kd_highSigma}
Kd \approx \frac{\pi}{1+\frac{2\lGC}{d}}, 
\end{equation} 
which is accurate within 1\,\% down to $d/\lGC \approx 7$. 

Eventually, when $d/\lGC \rightarrow \infty$, $Kd \rightarrow \pi$, and the potential and ion density profiles reach a limit, sometimes referred to as the ``Gouy-Chapman limit'': 
\begin{equation}
    \psi(z) = \ln\left\{\cos^2(\pi z/d)\right\}
\end{equation}

\begin{equation}
    n(z) = \dfrac{\pi}{2 d^2 \lB \cos^2(\pi z/d)}
\end{equation}

\subsection{General case: two walls with salt}
\label{sec:twowalls_salt}

\subsubsection{Exact solution} 

In the general case of an aqueous electrolyte confined between two symmetrical parallel walls located at $z=-d/2$ and $z=d/2$, the potential profile can be written in terms of the Jacobi elliptic functions cd, sn, cn and dn. 

As for the no co-ion regime (see Sec.~\ref{sec:twowalls_nosalt}), one can define an auxiliary reduced potential $\psi$, which is always negative: 
\begin{equation}
    \psi = -\text{sgn}(\Sigma) \times \phi = -\text{sgn}(\Sigma) \times \beta q V .
\end{equation}
The auxiliary potential writes: 
\begin{equation}
    \psi(y) = 2 \ln \left\{ \sqrt{m} \ \text{cd}\left( \frac{y r_1}{2 \sqrt{m}} , m^2 \right) \right\} , 
\end{equation}
where $y=z/d \in [-0.5;0.5]$ is the reduced position, $r_1 = d/\debye$, and $m=\exp(\psi_\text{m})$ with $\psi_\text{m}$ the potential in the middle of the slab.  

The parameter $m$ is controlled by $r_1$ and by the dimensionless surface charge $\debye/\lGC$ -- previously denoted as $\chi$, which we will denote $r_2$ in the following for consistency with $r_1$. Specifically, $m$ is the solution of: 
\begin{equation}
    r_2 = \frac{1-m^2}{2 \sqrt{m}} \times \frac{\text{sn}\left( \frac{r_1}{4 \sqrt{m}} , m^2 \right)}{\text{cn}\left( \frac{r_1}{4 \sqrt{m}} , m^2 \right) \text{dn}\left( \frac{r_1}{4 \sqrt{m}} , m^2 \right)} , 
\end{equation}
with $r_2 = \debye/\lGC$. 
For a given $r_1$, $m$ lies in the interval $]m_0,1[$, where $m_0$ the solution of $4\sqrt{m} K(m^2)=r_1$ (with $K(x)$ the complete elliptic integral of the first kind).

The PB equation can also be solved numerically (a Python script to that aim is provided in the appendix, \ref{sec:pb_slit_num}), and different approximate solutions for the auxiliary reduced potential $\psi$ expressed as a function of the reduced position $y=z/d$ can be written depending on $r_1$ and $r_2$, discussed in the following.

\subsubsection{Debye-H\"uckel limit} 
\label{sec:twowalls_salt_DH}

When the reduced potential $|\psi|$ is small everywhere (this condition will be quantified in section~\ref{sec:slit_r1r2}), the PB equation can be linearized (DH regime, already encountered when considering only one wall in Sec.~\ref{sec:onewall_salt}). 

With the conditions: 
\begin{equation*}
\left.\frac{\mathrm{d}\psi}{\mathrm{d}z}\right|_{z=0} = 0 
\quad \text{and} \quad 
\left.\frac{\mathrm{d}\psi}{\mathrm{d}z}\right|_{z=d/2} = -\frac{2}{\lGC}, 
\end{equation*}
the solution to the linearized PB equation, $\frac{\mathrm{d}^2 \psi}{\mathrm{d} z^2} = \psi / \debye^2$, writes: 
\begin{equation}
\psi 
= - \frac{2\debye}{\lGC} \ \frac{\cosh\left(\frac{z}{\debye}\right)}{\sinh\left(\frac{d}{2 \debye}\right)}
= -2 r_2 \ \frac{\cosh\left(y r_1\right)}{\sinh\left(\frac{r_1}{2}\right)} . 
\end{equation}
The corresponding potential $V$ writes: 
\begin{equation}
V (z)
= \frac{\Sigma \debye}{\eps} \ \frac{\cosh\left(\frac{z}{\debye}\right)}{\sinh\left(\frac{d}{2 \debye}\right)} .
\end{equation}
The limits of validity of this approximation will be discussed in section~\ref{sec:slit_r1r2}.

\subsubsection{Homogeneous/parabolic potential limits}
\label{sec:twowalls_IG}

\paragraph{Homogeneous potential} 

When the EDLs overlap ($r_1 = d/\debye \ll 1$) and for low enough surface charges ($r_1 r_2 = d/\lGC \ll 1$), the potential and ion densities are almost homogeneous: this is commonly referred to as the \emph{ideal gas} limit, and the potential identifies with the so-called Donnan potential \cite{Bocquet2010}. 
In this limit, the ion densities and the auxiliary reduced potential write \cite{Bocquet2010}: 
\begin{equation}\label{eq:n_overlap_smallSigma}
\begin{split}
    n^\text{IG}_\pm &= \mp \frac{\Sigma}{qd} + \sqrt{\ns^2 + \left(\frac{\Sigma}{qd}\right)^2} \\ 
    &= \ns \left\{ \mp \text{sgn}(\Sigma) \frac{\lDu}{d} + \sqrt{1+\left( \frac{\lDu}{d} \right)^2} \right\} , 
\end{split}
\end{equation}
\begin{equation}\label{eq:phi_overlap_smallSigma}
    \psi_\text{IG} = - \text{asinh} \left( \frac{\lDu}{d} \right) = - \text{asinh} \left( \frac{4r_2}{r_1} \right) ,  
\end{equation}
introducing the Dukhin length $\lDu = (|\Sigma|/q)/\ns = 4\debye^2/\lGC$. An example of potential profile in the ideal gas regime is shown in Fig.~\ref{fig:potential_2walls}. The limits of validity of this approximation will be discussed in section~\ref{sec:slit_r1r2}.

\bigskip
\paragraph{Parabolic potential} 

For weaker EDL overlaps, one can expand the solution of the PB equation in series of $z/\debye$; to second order, one obtains \cite{Silkina2019}: 
\begin{equation}
    \psi(z) \approx \psim + \frac{\text{sinh}(\psim)}{2} \left( \frac{z}{\debye} \right)^2 . 
\end{equation}
Using $\psim=\psi_\text{IG}$, one gets: 
\begin{equation}
    \psi(z) \approx - \text{asinh} \left( \frac{4r_2}{r_1} \right) - \frac{2 r_2}{r_1} \left( \frac{z}{\debye} \right)^2 . 
\end{equation}
However, the ideal gas potential is close to the average potential in the channel, and systematically shifted from the potential in the middle of the channel. The shift can be estimated as: 
\begin{equation*}
    \frac{2}{d} \int_0^{d/2} - \frac{2 r_2}{r_1} \left( \frac{z}{\debye} \right)^2 \dd z \approx -\frac{r_1 r_2}{6}, 
\end{equation*}
and it can be subtracted from $\psi_\text{IG}$ to obtain a closer approximation to $\psim$: 
\begin{equation}
    \psi_\text{IG}^\text{cor} = - \text{asinh} \left( \frac{4r_2}{r_1} \right) + \frac{r_1 r_2}{6}. 
\end{equation}
Overall, a good practical approximation for $\psi(z)$ is:
\begin{equation}\label{eq:psi_parabolic_practical}
\begin{split}
    \psi(z) &\approx \psi_\text{IG}^\text{cor} + \frac{\text{sinh}(\psi_\text{IG})}{2} \left( \frac{z}{\debye} \right)^2 \\
    &\approx - \text{asinh} \left( \frac{4r_2}{r_1} \right) + \frac{r_1 r_2}{6} - \frac{2 r_2}{r_1} \left( \frac{z}{\debye} \right)^2 \\
    &\approx - \text{asinh} \left( \frac{4r_2}{r_1} \right) + \frac{r_1 r_2}{6} - 2 r_1 r_2 y^2. 
\end{split}
\end{equation}
An example of potential profile in the parabolic potential regime is shown in Fig.~\ref{fig:potential_2walls}. The limits of validity of this approximation will be discussed in section~\ref{sec:slit_r1r2}.

\subsubsection{No co-ion limit} 
\label{sec:slit_r1r2_nocoion}

When the EDLs overlap and for large enough surface charges, co-ions are excluded from the channel and the PB equation can be solved with counter-ions only, see Sec.~\ref{sec:twowalls_nosalt}. 
The auxiliary reduced potential $\psi(r)$ and the counter-ion density profile $n(r)$ were given by Eq.~\eqref{eq:psi_twowalls_nosalt} and by Eq.~\eqref{eq:n_twowalls_nosalt}, respectively. 
Both $\psi$ and $n$ depended on an inverse length $K$ related to the surface charge via Eq.~\eqref{eq:Kd_vs_doverlGC}. 

However, the potential in the middle of the channel was arbitrarily fixed to zero. 
To reproduce the exact potential,  
one needs to acknowledge that the channel is connected to an external salt reservoir with salt concentration $n_0$ and corresponding Debye length $\debye$.  
Moving the reference of potential to the reservoir, $\psi_0=0$, the Boltzmann distribution for the counter-ions becomes $n(z) = \ns\,\mathrm{e}^{-\psi(z)}$, and the PB equation rewrites: 
\begin{equation*}
    \dfrac{\mathrm{d}^2 \psi}{\mathrm{d} z^2} = -\frac{1}{2\debye^2} \,\mathrm{e}^{-\psi(z)}, 
\end{equation*}
\begin{equation*}
    \text{with}\ \ \left.\frac{\dd \psi}{\dd z}\right|_{z=0} = 0 
    \ \ \text{and}\ \ \left.\frac{\dd \psi}{\dd z}\right|_{z=d/2} = -\frac{2}{\lGC}. 
\end{equation*}
Denoting $Kd = r_3$, the solution writes: 
\begin{equation}\label{eq:nocoion_limit}
    \psi = 2\ln \left[ \frac{r_1}{2r_3} \cos(r_3 y) \right] , 
\end{equation}
with 
\begin{equation}\label{eq:nocoion_ter3}
r_3 \,\tan\left(\frac{r_3}{2}\right) = r_1 r_2.
\end{equation} 
In particular, the potential in the center of the channel is $\psim = \psi(y=0) = 2\ln \left[ r_1/(2r_3) \right]$. 
An example of potential profile in the no co-ion regime is shown in Fig.~\ref{fig:potential_2walls}. 
The limits of validity of this approximation will be discussed in section~\ref{sec:slit_r1r2}.

\subsection{Validity of the approximate solutions}
\label{sec:slit_r1r2}

\begin{figure}
    \centering
    \includegraphics[width=\linewidth]{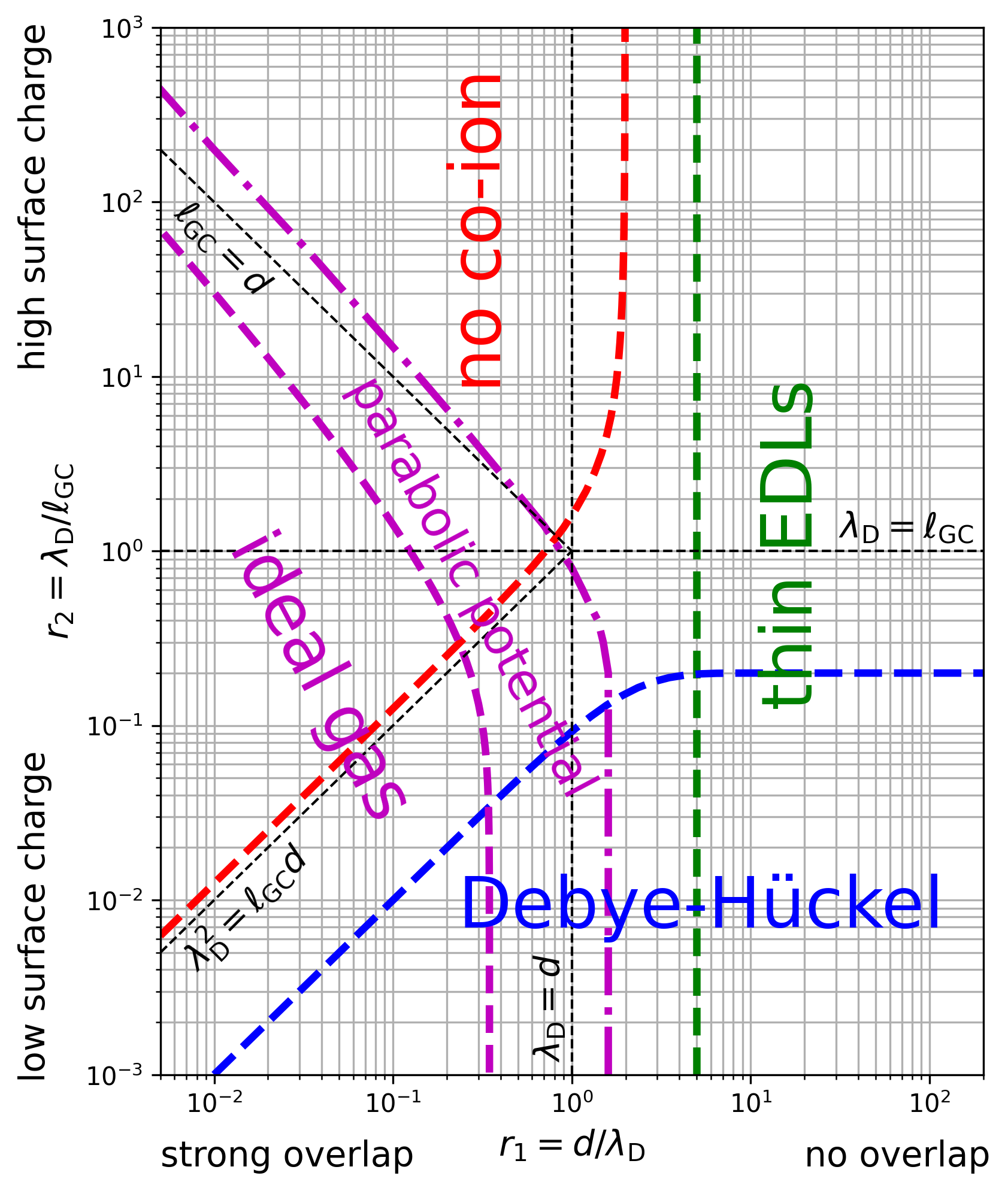}
    \caption{Domains of validity of the different approximate solutions to the PB equation for an electrolyte solution confined between two planar walls, as a function of the ratios $r_1 = d/\debye$ and $r_2 = \debye/\lGC$, where $d$ is the distance between the walls, $\debye$ the Debye length, and $\lGC$ the Gouy-Chapman length.}
    \label{fig:r1r2map}
\end{figure}

\begin{figure}
    \centering
    \includegraphics[width=0.8\linewidth]{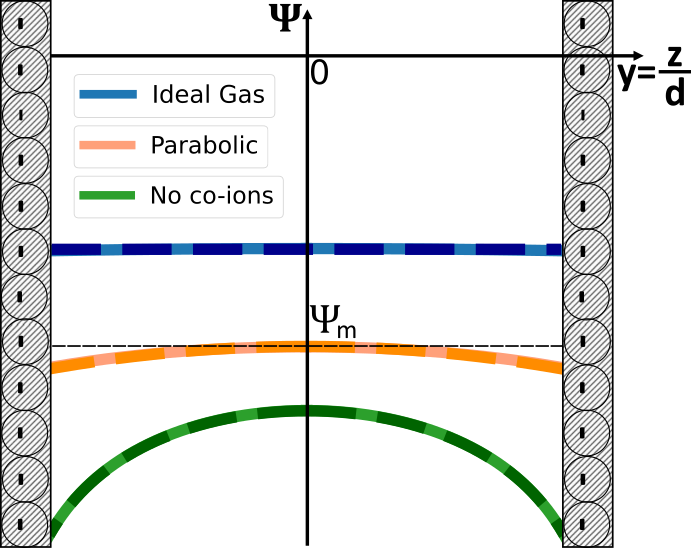}
    \caption{Typical potential profiles for an electrolyte solution confined between two planar walls, in the strong overlap regime ($r_1=d/\debye = 0.1$). Three limiting regimes are illustrated: ideal gas/homogeneous potential ($r_2=\debye/\lGC=1$), parabolic potential ($r_2=10$), and no co-ion ($r_2=100$) limits. In the three cases, the exact potential (light continuous line) is compared to the approximated formula (dark dashed line) for the ideal gas, Eq.~\eqref{eq:phi_overlap_smallSigma}, parabolic potential, Eq.~\eqref{eq:psi_parabolic_practical}, and no co-ion, Eq.~\eqref{eq:nocoion_limit}, limits.}
    \label{fig:potential_2walls}
\end{figure}

In section~\ref{sec:twowalls_salt}, it appeared that the potential profile is uniquely determined by the two ratios $r_1 = d/\debye$ and $r_2 = \debye/\lGC$. 
Here we will discuss the range of $r_1$ and $r_2$ where the different approximate solutions presented in the previous sections can safely be used. Following \citet{Markovich2016a}, we will illustrate the domains of validity of the approximate solutions in a $r_2 - r_1$ diagram, see Fig.~\ref{fig:r1r2map}. Typical potential profiles in different regimes corresponding to a strong EDL overlap are represented in Fig.~\ref{fig:potential_2walls}, where the exact solution is compared to the approximated formulas.

\subsubsection{Thin EDLs limit} 

When the distance $d$ between the surfaces is much larger than the Debye length (corresponding to well-separated EDLs), one can solve the PB equation for a single charged wall (see Sec.~\ref{sec:onewall_salt}), and superimpose the potentials of the two walls to obtain the full potential in the channel. 

In practice, this approach provides an an accurate representation of the exact solution as long as $r_1 = d/\debye > 5$. This boundary is represented with a green dashed line in Fig.~\ref{fig:r1r2map}.

\subsubsection{Debye-H\"uckel limit} 

In practice, the DH potential introduced in Sec.~\ref{sec:twowalls_salt_DH} is an excellent approximation to the exact potential as long as the surface potential, $|\psi_\text{s}| = \max(|\psi|) = |\psi(y=1/2)|$, remains under a (somehow arbitrary) critical value $\psi_\text{c} = 0.4$. This corresponds to $r_2 \leq \tanh(r_1/2) \times \psi_\text{c}/2$. This boundary is represented with a blue dashed line in Fig.~\ref{fig:r1r2map}. 

One can then simplify the relation between $r_1$ and $r_2$ by considering two limits: for thin EDLs, $r_1 \gg 1$, the DH solution can be used up to $r_2 \leq \psi_\text{c}/2$; for overlapping EDLs, $r_1 \ll 1$, the DH solution can be used up to $r_2 \leq r_1 \times \psi_\text{c}/4$.

\subsubsection{Homogeneous/parabolic potential limits} 

\paragraph{Homogeneous potential (ideal gas)}

To estimate the range of validity of the homogeneous potential approximation (see Sec.~\ref{sec:twowalls_IG}), one can compare the homogeneous potential value to the next term in a Taylor series of the potential as a function of $z/\debye$, i.e., the parabolic term. 

The homogeneous potential writes: $|\psi_\text{IG}| = \text{asinh}(4r_2/r_1)$, and the maximum difference between the parabolic term and its average value writes: $r_1 r_2 /3$. Imposing a maximum ratio $\theta_\text{c}$ between the two terms, one obtains: 
\begin{equation}
    r_1 r_2 = 3\theta_\text{c} \text{asinh} \left( \frac{4 r_2}{r_1} \right) . 
\end{equation}
This boundary is represented for $\theta_\text{c} = 1\,\%$ with a purple dashed line in Fig.~\ref{fig:r1r2map}. 

\bigskip
\paragraph{Parabolic potential}

The accuracy of the parabolic potential approximation depends crucially on the choice of $\psim$. Using $\psim=\psi_\text{IG}$, the error is approximately half the one of the homogeneous potential approximation, which does not extend much the range of parameters where it can be applied.  
In contrast, Eq.~\eqref{eq:psi_parabolic_practical} approximates the exact potential within 1\,\% over a broader range of parameters, see the purple dashed-dotted line in Fig.~\ref{fig:r1r2map}.

\subsubsection{No co-ion limit} 

To determine the limits of applicability of this regime, one can estimate the ratio $\eta$ between counter-ions and co-ion densities in the middle of the channel: 
\begin{equation}\label{eq:nocoion_R}
    \eta = \frac{\nm^\text{counter-ion}}{\nm^\text{co-ion}} = \exp{-2\psim} = \left(\frac{2 r_3}{r_1}\right)^4 .
\end{equation}  
One can consider that co-ions are efficiently excluded from the channel when $\eta$ is above a critical value $\eta_\text{c}$, that we will arbitrarily fix to $100$. 
Combining Eq.~\eqref{eq:nocoion_ter3} and Eq.~\eqref{eq:nocoion_R}, one can see that $\eta=\eta_\text{c}$ corresponds to: 
\begin{equation}
r_2 = \frac{\eta_\text{c}^{1/4}}{2} \,\tan\left(\frac{r_1 \eta_\text{c}^{1/4}}{4}\right).
\end{equation}
This boundary is represented with a red dashed line in Fig.~\ref{fig:r1r2map}. 

One can simplify the relation between $r_1$ and $r_2$ by considering two limits: for small surface charges, $r_2 \ll 1$, the no co-ion approximation can be used for $r_2 \geq r_1 \eta_\text{c}^{1/2}/8 \approx 1.25 r_1$; for large surface charges, $r_2 \gg 1$, the no co-ion approximation can be used for $r_1 \leq 2\pi / \eta_\text{c}^{1/4} \approx 1.99$.

\section{Cylindrical channel}
\label{sec:cyl}

This section gathers the equations describing a $Z$:$Z$ electrolyte solution confined inside a cylindrical channel of radius $R$, see Fig.~\ref{fig:PB_figs}(d). More detailed discussions can be found in e.g. Refs.~\citenum{Rice1985,Berg2009,Berg2011}.

\subsection{Thin EDL}
\label{sec:cyl_thin_edl}

When the channel radius $R$ is much larger than the thickness of the EDL, 
one can neglect the wall curvature at the scale of the EDL, and solve the PB equation for a single planar charged wall, see section~\ref{sec:onewall_salt}, where the distance to the wall is $z=R-r$. 

Accordingly, the integrals computed per unit surface for a single planar wall in Sec.~\ref{sec:onewall_salt_integrals} can simply be multiplied by $2\pi R$ to obtain the corresponding integral per unit length of the cylindrical channel. 

The limits of validity of this approximation will be quantified in section~\ref{sec:cyl_limits}.

\subsection{Thick EDL, high surface charge: no co-ion}
\label{sec:cyl_nosalt}

When the EDLs overlap and for large enough surface charges (these conditions will be quantified in section~\ref{sec:cyl_limits}), co-ions are excluded from the channel and the PB equation can be solved with counter-ions only.

\subsubsection{PB equation: derivation}

As for the slit channel, the counter-ion density is denoted $n(r)$, and one can define an auxiliary reduced potential $\psi$, which is always negative: 
\begin{equation*}
    \psi = -\text{sgn}(\Sigma) \times \phi = -\text{sgn}(\Sigma) \times \beta q V .
\end{equation*}
The potential in the middle of the channel is arbitrarily fixed to zero: $\psim = \psi(r=0) = 0$ (this condition will be relaxed in section~\ref{sec:cyl_r1r2_nocoion}).  
Denoting $\nm$ the counter-ion density at $r=0$, one can define the same characteristic length $K^{-1}$  as for the slit geometry,
\begin{equation*}
    K^{-1} = \frac{1}{\sqrt{2 \pi \lB \nm}} 
    \quad \Leftrightarrow \quad 
    \nm = \frac{K^2}{2\pi \lB}, 
\end{equation*}
and derive the PB equation for a negative surface charge (with positive counter-ions) or for a positive surface charge (with negative counter-ions).  

\bigskip
\paragraph*{Poisson equation:}
\begin{equation*}
    q n(r) = -\dfrac{\eps}{\beta q} 
    \,\frac{1}{r} \frac{\mathrm{d}}{\mathrm{d} r} \left( r \frac{\mathrm{d} \psi}{\mathrm{d} r}\right)
\end{equation*}

\paragraph*{Boltzmann distribution:}
\begin{equation*}
    n(r) = \nm\,\mathrm{e}^{-\psi(r)}
\end{equation*}

\paragraph*{Poisson-Boltzmann equation:}
\begin{equation}
    \frac{1}{r} \frac{\mathrm{d}}{\mathrm{d} r} \left( r \frac{\mathrm{d} \psi}{\mathrm{d} r}\right) = -2 K^2 \,\mathrm{e}^{-\psi(r)}
\end{equation}

\subsubsection{PB equation: solution} 

\begin{equation}\label{eq:psi_cyl_nosalt}
    \psi (r) = 2 \ln\left[1-\left(\frac{Kr}{2}\right)^2\right]
\end{equation}
\begin{equation}\label{eq:dr_psi}
    \dfrac{\mathrm{d}\psi}{\mathrm{d}r} = \frac{-K^2 r}{1-\left(\frac{Kr}{2}\right)^2}
\end{equation}
\begin{equation}\label{eq:n_cyl_nosalt}
    n (r) = \dfrac{\nm}{\left[1-\left(\frac{Kr}{2}\right)^2\right]^2}  = \dfrac{K^2}{2 \pi \lB \left[1-\left(\frac{Kr}{2}\right)^2\right]^2}
\end{equation}

\subsubsection{Surface charge and surface potential}

To fully determine the potential and ion density profiles, one needs to express $K$ as a function of the surface charge. 

Electrostatics at the interface (or equivalently, global electroneutrality): 
\begin{equation}
\left.\frac{\mathrm{d}V}{\mathrm{d}r}\right|_{r=R} = \frac{\Sigma}{\eps} 
\quad \Rightarrow \quad 
\left.\frac{\mathrm{d}\psi}{\mathrm{d}r}\right|_{r=R} = - \frac{2}{\lGC}
\end{equation}

Using Eq.~\eqref{eq:dr_psi}, one obtains: 
\begin{equation}\label{eq:KR_vs_RoverlGC}
\frac{(KR)^2}{1-\left(\frac{KR}{2}\right)^2} = \frac{2R}{\lGC} , 
\end{equation} 
which can be solved: 
\begin{equation}\label{eq:KR_vs_RoverlGC2}
\frac{KR}{2} = \sqrt{\frac{R/\lGC}{R/\lGC+2}} . 
\end{equation} 
Hence, in contrast with the slit geometry, an explicit expression of $K$ can be derived in general for a cylindrical channel.

\subsubsection{A few integrals}

\paragraph{Electrostatic energy $\mathcal{E}$ (per unit length of the channel)}

Denoting $y = KR/2$, 
\begin{multline}\label{eq:E_cyl_nocoion}
\mathcal{E} 
= \frac{\eps}{2} \int_0^R 2\pi r \,\mathrm{d}r\ \left( \frac{\mathrm{d}V}{\mathrm{d}r} \right)^2 \\ 
= \frac{2}{\lB \beta} \left[ \ln (1-y^2) + \frac{y^2}{1-y^2} \right] 
\end{multline} 

Using Eq.~\eqref{eq:KR_vs_RoverlGC2}, $\mathcal{E}$ can be directly written in terms of $R/\lGC$: 
\begin{equation}\label{eq:E_cyl_nocoion2}
\mathcal{E} 
= \frac{2}{\lB \beta} \left[ \ln (\frac{2}{R/\lGC+2}) + \frac{R}{2\lGC} \right] 
\end{equation}

\paragraph{Ionic density} 

\begin{equation}
\int_0^R 2\pi r \,\mathrm{d}r\  n(r)  
= 2\pi R \frac{|\Sigma|}{q} = \frac{R}{\lB \lGC} 
\end{equation}

\subsubsection{Low and high surface charge limits}

\paragraph{Low surface charge}

When $\Sigma$ is small enough that $R/\lGC \ll 1$, Eq.~\eqref{eq:KR_vs_RoverlGC2} simplifies: 
\begin{equation}\label{eq:KR_vs_RoverlGC_IG}
KR \approx \sqrt{2R/\lGC},
\end{equation} 
which is accurate to within 1\,\% up to $R/\lGC\approx 0.04$.

Note that when $R/\lGC \ll 1$, $KR \ll 1$, so that $\psi \approx 0$ and the ion density is approximately homogeneous in the channel: 
\begin{equation}
    n \approx \dfrac{K^2}{2 \pi \lB} = \dfrac{1}{\pi R \lB \lGC}.
\end{equation}
This is commonly called the \emph{ideal gas} regime. In practice, the ion density varies by less than 10\,\% over the channel thickness as long as $R/\lGC < 0.1$.

\bigskip
\paragraph{High surface charge}

When $\Sigma$ is large enough that $R/\lGC \gg 1$, Eq.~\eqref{eq:KR_vs_RoverlGC2} simplifies:
\begin{equation}\label{eq:KR_vs_RoverlGC_GC}
KR \approx \frac{2}{1+\frac{\lGC}{R}},
\end{equation} 
which is accurate to within 1\,\% down to $R/\lGC \approx 7$.

Eventually, when $R/\lGC \rightarrow \infty$, $KR \rightarrow 2$, and the potential and ion density profiles reach a limit, sometimes referred to as the ``Gouy-Chapman limit'':  
\begin{equation}
    \psi (r) = 2\ln\left[1-\left(\frac{r}{R} \right)^2 \right], 
\end{equation}
\begin{equation}
    n (r) = \dfrac{2}{\pi \lB R^2 \left[ 1 - \left( \frac{r}{R} \right)^2 \right]^2}.
\end{equation}

\subsection{General case}
\label{sec:cyl_salt}

In a cylindrical channel, we are not aware of a usable analytical solution for the general case, and the PB equation has to be solved numerically (a Python script to that aim is provided in the appendix, \ref{sec:pb_cyl_num}). 
Nevertheless, like in the slit case, one can define two ratios $r_1 = R/\debye$ and $r_2 = \debye/\lGC$, which uniquely determine the auxiliary reduced potential $\psi = -\text{sgn}(\Sigma) \times \phi$ expressed as a function of the reduced position $y=r/R$. Different approximate solutions can be written depending on $r_1$ and $r_2$, discussed in the following.

\subsubsection{Debye-H\"uckel limit} 
\label{sec:cyl_salt_DH}

When the reduced potential $|\psi|$ is small everywhere, the PB equation can be linearized (DH regime).

With the conditions: 
\begin{equation*}
\left.\frac{\mathrm{d}\psi}{\mathrm{d}r}\right|_{r=0} = 0 
\quad \text{and} \quad 
\left.\frac{\mathrm{d}\psi}{\mathrm{d}r}\right|_{r=R} = -\frac{2}{\lGC}, 
\end{equation*}
the solution to the linearized PB equation in cylindrical coordinates, $\frac{1}{r} \frac{\mathrm{d}}{\mathrm{d} r} \left( r \frac{\mathrm{d} \psi}{\mathrm{d} r}\right) = \psi / \debye^2$, writes: 
\begin{equation}
\psi (r)
= - \frac{2\debye}{\lGC} \ \frac{\mathrm{I}_0 \left(\frac{r}{\debye}\right)}{\mathrm{I}_1 \left(\frac{R}{\debye}\right)} = - 2r_2 \frac{\mathrm{I}_0 \left(r_1 y\right)}{\mathrm{I}_1 \left(r_1\right)}, 
\end{equation}
where $\mathrm{I}_0 (x)$ and $\mathrm{I}_1 (x)$ are the modified Bessel functions of the first kind of order zero and one. 
The corresponding potential $V$ writes: 
\begin{equation}
V (r)
= \frac{\Sigma \debye}{\eps} \ \frac{\mathrm{I}_0 \left(\frac{r}{\debye}\right)}{\mathrm{I}_1 \left(\frac{R}{\debye}\right)} .
\end{equation}

The limits of validity of this approximation will be quantified in section~\ref{sec:cyl_limits}.

\bigskip
\paragraph*{Electrostatic energy $\mathcal{E}$ (per unit length of the cylindrical channel)}

With $y = R/\debye$, 
\begin{equation}
\begin{split}
    \mathcal{E} &= \frac{\eps}{2} \int_0^R 2 \pi r \left( \frac{\dd V}{\dd r}\right)^2 \dd r \\
    &= \frac{R^2}{2\beta \lB \lGC^{\,2}} \left[ 1 - \frac{I_0(y) I_2(y)}{I_1(y)^2} \right].
\end{split}
\end{equation}

\subsubsection{Homogeneous/parabolic potential limits}
\label{sec:cyl_IG}

\paragraph{Homogeneous potential} 
 
When the EDLs overlap ($r_1 = R/\debye \ll 1$) and for low enough surface charges ($r_1 r_2 = R/\lGC \ll 1$), the potential and ion densities are almost homogeneous: this is commonly referred to as the \emph{ideal gas} limit, and the potential identifies with the so-called Donnan potential \cite{Bocquet2010}. 
In this limit, the ion densities and the auxiliary reduced potential write: 
\begin{equation}\label{eq:n_overlap_smallSigma_cyl}
\begin{split}
    n^\text{IG}_\pm &= \mp \frac{\Sigma}{qR} + \sqrt{\ns^2 + \left(\frac{\Sigma}{qR}\right)^2} \\ 
    &= \ns \left\{ \mp \text{sgn}(\Sigma) \frac{\lDu}{R} + \sqrt{1+\left( \frac{\lDu}{R} \right)^2} \right\} , 
\end{split}
\end{equation}
\begin{equation}\label{eq:phi_overlap_smallSigma_cyl}
    \psi_\text{IG} = - \text{asinh} \left( \frac{\lDu}{R} \right) = - \text{asinh} \left( \frac{4r_2}{r_1} \right) ,  
\end{equation}
introducing the Dukhin length $\lDu = (|\Sigma|/q)/\ns = 4\debye^2/\lGC$. 
The limits of validity of this approximation will be discussed in section~\ref{sec:cyl_limits}. 

\bigskip
\paragraph{Parabolic potential} 

For weaker EDL overlaps, one can expand the solution of the PB equation in series of $r/\debye$; to second order, one obtains \cite{Silkina2019}: 
\begin{equation}
    \psi(r) \approx \psim + \frac{\text{sinh}(\psim)}{4} \left( \frac{r}{\debye} \right)^2 . 
\end{equation}
Using $\psim=\psi_\text{IG}$, one gets: 
\begin{equation}
    \psi(r) \approx - \text{asinh} \left( \frac{4r_2}{r_1} \right) - \frac{r_2}{r_1} \left( \frac{r}{\debye} \right)^2 . 
\end{equation}
However, the ideal gas potential is close to the average potential in the channel, and systematically shifted from the potential in the middle of the channel. 
The shift can be estimated as: 
\begin{equation*}
    \frac{1}{\pi R^2} \int_0^{R} - \frac{r_2}{r_1} \left( \frac{r}{\debye} \right)^2 2\pi r \dd r \approx -\frac{r_1 r_2}{2}, 
\end{equation*}
and it can be subtracted from $\psi_\text{IG}$ to obtain a closer approximation to $\psim$: 
\begin{equation}
    \psi_\text{IG}^\text{cor} = - \text{asinh} \left( \frac{4r_2}{r_1} \right) + \frac{r_1 r_2}{2}. 
\end{equation}
Overall, a good practical approximation for $\psi(r)$ is:
\begin{equation}\label{eq:psi_parabolic_practical_cyl}
\begin{split}
    \psi(r) &\approx \psi_\text{IG}^\text{cor} + \frac{\text{sinh}(\psi_\text{IG})}{4} \left( \frac{r}{\debye} \right)^2 \\
    &= - \text{asinh} \left( \frac{4r_2}{r_1} \right) + \frac{r_1 r_2}{2} - \frac{r_2}{r_1} \left( \frac{r}{\debye} \right)^2 . 
\end{split}
\end{equation}
The limits of validity of this approximation will be discussed in section~\ref{sec:cyl_limits}.

\subsubsection{No co-ion limit} 
\label{sec:cyl_r1r2_nocoion}

When the EDL extends over the whole channel and for large enough surface charges, co-ions are excluded from the channel and the PB equation can be solved with counter-ions only, see Sec.~\ref{sec:cyl_nosalt}. The auxiliary reduced potential $\psi(r)$ and the counter-ion density profile $n(r)$ were given by Eq.~\eqref{eq:psi_cyl_nosalt} and by Eq.~\eqref{eq:n_cyl_nosalt}, respectively. Both $\psi$ and $n$ depended on an inverse length $K$ related to the surface charge via Eq.~\eqref{eq:KR_vs_RoverlGC2}. 

However, the potential in the middle of the channel was arbitrarily fixed to zero. To reproduce the exact potential,  
one needs to acknowledge that the channel is connected to an external salt reservoir with salt concentration $n_0$ and corresponding Debye length $\debye$. 
Moving the reference of potential to the reservoir, $\psi_0=0$, the Boltzmann distribution for the counter-ions becomes $n(z) = \ns\,\mathrm{e}^{-\psi(z)}$, and the PB equation rewrites: 
\begin{equation*}
        \frac{1}{r} \frac{\mathrm{d}}{\mathrm{d} r} \left( r \frac{\mathrm{d} \psi}{\mathrm{d} r}\right) = -\frac{1}{2\debye^2} \,\mathrm{e}^{-\psi(r)} , 
\end{equation*}
\begin{equation*}
    \text{with}\ \ \left.\frac{\mathrm{d}\psi}{\mathrm{d}r}\right|_{r=0} = 0 
    \ \ \text{and}\ \ \left.\frac{\mathrm{d}\psi}{\mathrm{d}r}\right|_{r=R} = -\frac{2}{\lGC}. 
\end{equation*}
Denoting $KR = r_3$, the solution writes: 
\begin{equation}\label{eq:nocoion_limit_cyl}
    \psi = 2\ln \left\{ \frac{r_1}{2r_3} \left[1-\left(\frac{r_3 y}{2}\right)^2\right] \right\} , 
\end{equation}
with 
\begin{equation}\label{eq:cyl_nocoion_ter3}
r_3 = 2 \sqrt{\frac{r_1 r_2}{r_1 r_2+2}} .
\end{equation} 
In particular, the potential in the center of the channel is $\psim = \psi(y=0) = 2\ln \left[ r_1/(2r_3) \right]$. 
The limits of validity of this approximation will be discussed in section~\ref{sec:cyl_limits}.

\subsection{Validity of the approximate solutions}
\label{sec:cyl_limits}

\begin{figure}
    \centering
    \includegraphics[width=\linewidth]{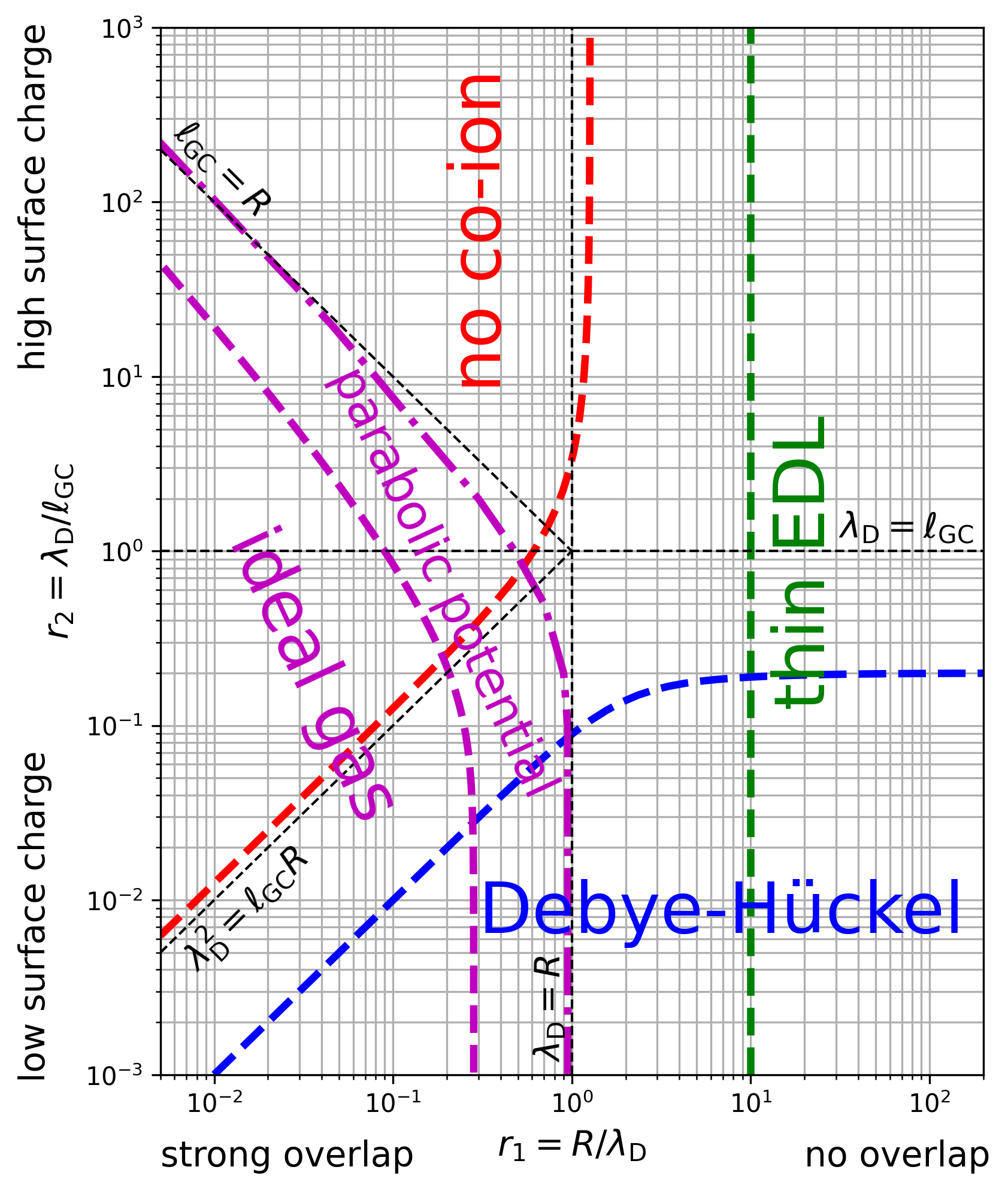}
    \caption{Domains of validity of the different approximate solutions to the PB equation for an electrolyte solution confined inside a cylindrical channel, as a function of the ratios $r_1 = R/\debye$ and $r_2 = \debye/\lGC$, where $R$ is the channel radius, $\debye$ the Debye length, and $\lGC$ the Gouy-Chapman length.}
    \label{fig:cyl_r1r2map}
\end{figure}

We will in the following bound the regions in the $r_1-r_2$ diagram where the different approximate solutions can be used safely, see Fig.~\ref{fig:cyl_r1r2map}. 

\subsubsection{Thin EDL limit} 

When the channel radius $R$ is much larger than $\debye$, one can neglect the wall curvature at the scale of the EDL, and solve the PB equation for a single planar charged wall. 

However this approximation should be taken with caution. Even when the EDL is significantly thinner than the channel radius ($R/\debye>10$), the planar wall solution only provides a fair description of the exact potential profile. Still, this boundary is represented with a green dashed line in Fig.~\ref{fig:r1r2map}.

\subsubsection{Debye-H\"uckel limit} 

Like in the slit case, the DH potential introduced in Sec.~\ref{sec:cyl_salt_DH} is an excellent approximation to the exact potential as long as the surface potential, $|\psi_\text{s}| = \max(|\psi|) = |\psi(r=R)|$ remains under a (somehow arbitrary) critical value $\psi_\text{c} = 0.4$. This corresponds to:
\begin{equation}
    r_2 \leq \frac{\mathrm{I}_1(r_1)}{\mathrm{I}_0(r_1)} \times \frac{\psi_\text{c}}{2}.
\end{equation}
This boundary is represented with a blue dashed line in Fig.~\ref{fig:cyl_r1r2map}. 

One can then simplify the relation between $r_1$ and $r_2$ by considering two limits: for thin EDLs, $r_1 \gg 1$, the DH solution can be used up to $r_2 \leq \psi_\text{c}/2$; for overlapping EDLs, $r_1 \ll 1$, the DH solution can be used up to $r_2 \leq r_1 \times \psi_\text{c}/4$.

\subsubsection{Homogeneous/parabolic potential limits} 

\paragraph{Homogeneous potential (ideal gas)}

To estimate the range of validity of the homogeneous potential approximation (see Sec.~\ref{sec:cyl_IG}), one can compare the homogeneous potential value to the next term in a Taylor series of the potential as a function of $R/\debye$, i.e., the parabolic term. 

The homogeneous potential writes: $|\psi_\text{IG}| = \text{asinh}(4r_2/r_1)$, and the maximum difference between the parabolic term and its average value writes: $r_1 r_2 /2$. Imposing a maximum ratio $\theta_\text{c}$ between the two terms, one obtains: 
\begin{equation}
    r_1 r_2 = 2\theta_\text{c} \text{asinh} \left( \frac{4 r_2}{r_1} \right) . 
\end{equation}
This boundary is represented for $\theta_\text{c} = 1\,\%$ with a purple dashed line in Fig.~\ref{fig:cyl_r1r2map}. 

\bigskip
\paragraph{Parabolic potential}

The accuracy of the parabolic potential approximation depends crucially on the choice of $\psim$. Using $\psim=\psi_\text{IG}$, the error is approximately the same as the one of the homogeneous potential approximation. 
In contrast, Eq.~\eqref{eq:psi_parabolic_practical_cyl} approximates the exact potential within 1\,\% over a broader range of parameters, see the purple dashed-dotted line in Fig.~\ref{fig:cyl_r1r2map}.

\subsubsection{No co-ion limit} 

To determine the limits of applicability of this regime, one can estimate the ratio $\eta$ between counter-ions and co-ion densities in the middle of the channel: 
\begin{equation}\label{eq:cyl_nocoion_R}
    \eta = \frac{\nm^\text{counter-ion}}{\nm^\text{co-ion}} = \exp{-2\psim} = \left(\frac{2 r_3}{r_1}\right)^4 .
\end{equation}  
One can consider that co-ions are efficiently excluded from the channel when $\eta$ is above a critical value $\eta_\text{c}$, that we will arbitrarily fix to $100$. 
Combining Eq.~\eqref{eq:cyl_nocoion_ter3} and Eq.~\eqref{eq:cyl_nocoion_R}, one can see that $\eta=\eta_\text{c}$ corresponds to: 
\begin{equation}
r_2 = \frac{2 r_1 \eta_\text{c}^{1/2}}{16 - r_1^{\,2} \eta_\text{c}^{1/2}} .
\end{equation}
This boundary is represented with a red dashed line in Fig.~\ref{fig:cyl_r1r2map}. 

One can simplify the relation between $r_1$ and $r_2$ by considering two limits: for small surface charges, $r_2 \ll 1$, the no co-ion approximation can be used for $r_2 \geq r_1 \eta_\text{c}^{1/2}/8 \approx 1.25 r_1$; for large surface charges, $r_2 \gg 1$, the no co-ion approximation can be used for $r_1 \leq 4 / \eta_\text{c}^{1/4} \approx 1.26$.

\section{Common boundary conditions}
\label{sec:BC}

\subsection{Surface charge density and surface potential}
\label{sec:interfacial_ES}

At a charged surface, the surface charge density $\Sigma$ and the surface potential $\phis$ can be linked by expressing the electric field at the wall as a function of the surface charge, Eq.~\eqref{eq:interface_ES_onewall}. This equation can be solved when the functional form of the potential is known, resulting e.g. in the Grahame equation, Eq.~\eqref{eq:grahame_onewall}, for one wall with salt. 

\begin{figure}
    \centering
    \includegraphics[width=\linewidth]{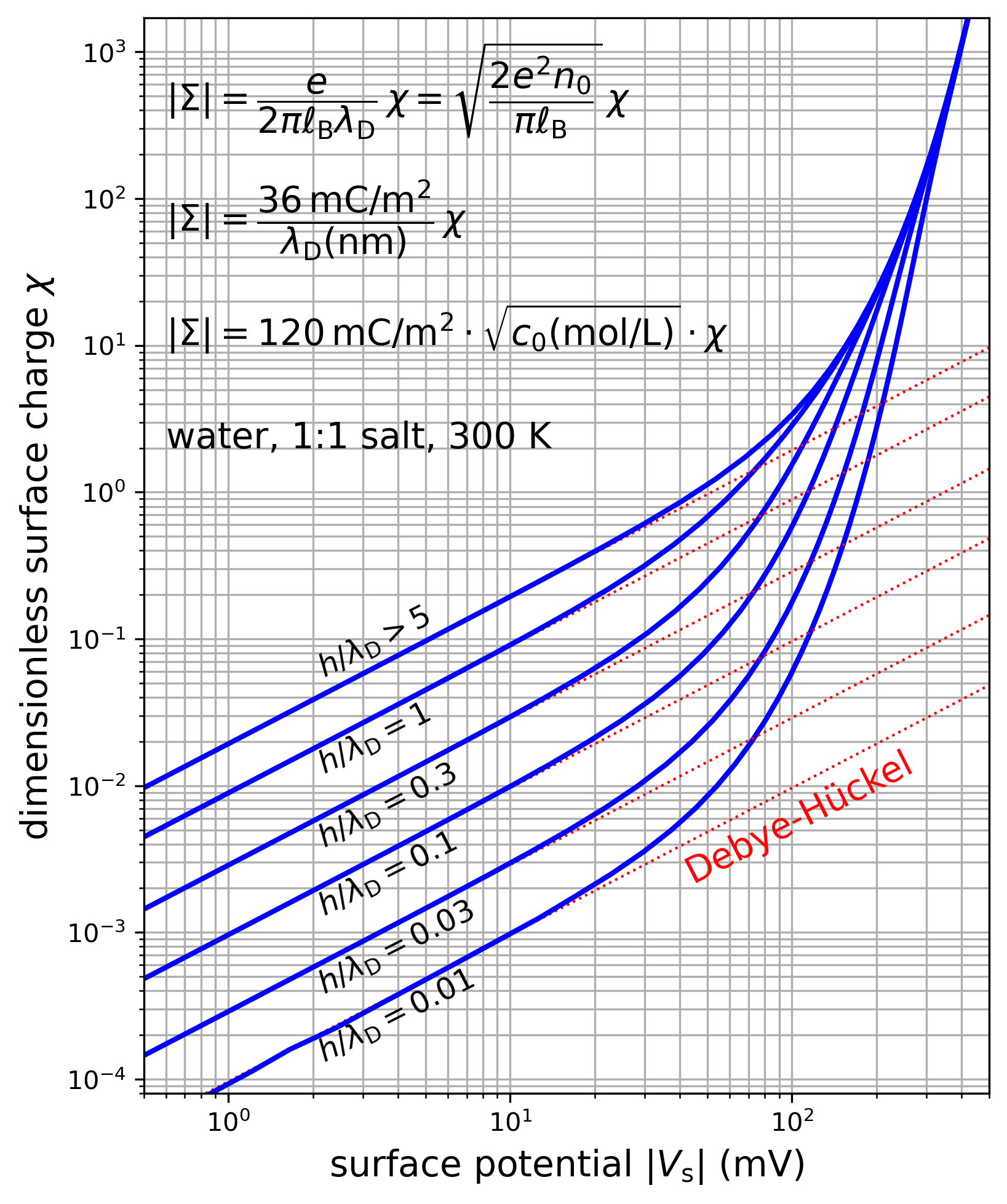}
    \caption{Dimensionless surface charge versus surface potential for an aqueous 1:1 electrolyte confined between two parallel walls, at 300\,K. Exact solution (blue lines) and Debye-H\"uckel approximation (red dotted lines) for different $h/\debye$ ratios.} 
    \label{fig:Sigma-Vs}
\end{figure}

In the general case of two walls with salt, one can solve the PB equation numerically to plot the relation between the surface charge and the surface potential, see Fig.~\ref{fig:Sigma-Vs}.

\subsection{Constant charge}

Insulating walls are commonly described using a constant surface charge, i.e., the surface charge density $\Sigma$ is imposed. This is the boundary condition considered by default in this formulary. The surface potential $\phis$ can then be obtained from interfacial electrostatics, see Sec.~\ref{sec:interfacial_ES}.

\subsection{Constant potential}

In contrast, conducting surfaces are usually described using a constant potential, i.e., the surface potential $\phis$ is imposed. 
All the results in the formulary can be expressed in terms of the surface potential $\phis$ and/or $\gamma = \tanh(\phis/4)$, through the relation between $\phis$ and $\Sigma$ given by interfacial electrostatics, see Sec.~\ref{sec:interfacial_ES}. 

\begin{figure}
    \centering
    \includegraphics[width=\linewidth]{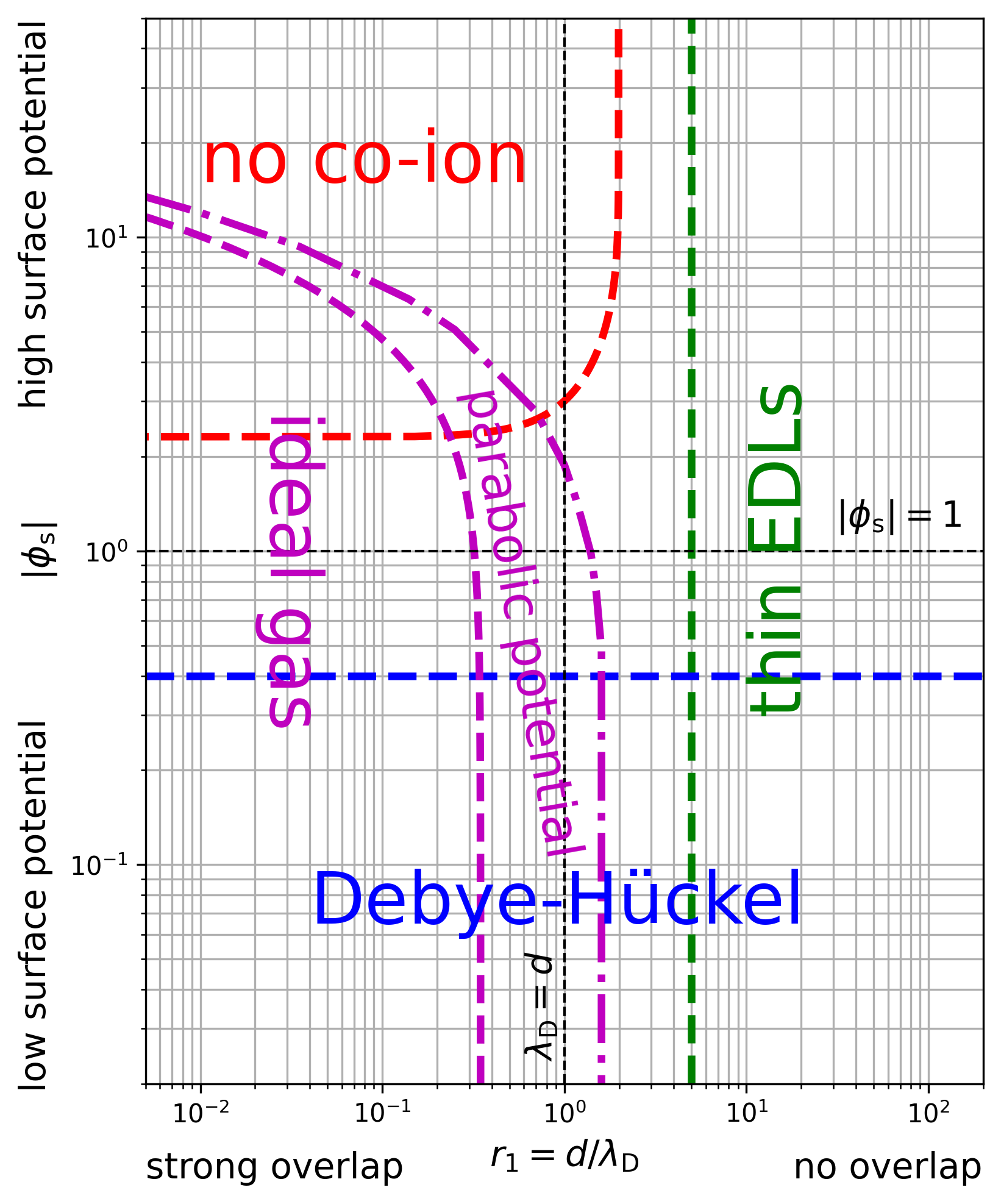}
    \caption{Domains of validity of the different approximate solutions to the PB equation for an electrolyte solution confined between two planar walls, as a function of $r_1 = d/\debye$ and $|\phis|$, where $d$ is the distance between the walls, $\debye$ the Debye length, and $\phis$ the reduced surface potential.}
    \label{fig:r1phismap}
\end{figure}

The domains of validity of the different approximations can also be represented, e.g. for a slit channel, as a function of $r_1 = d/\debye$ and $|\phis|$, see Fig.~\ref{fig:r1phismap}. 

\subsection{Charge regulation} 

Finally, surface charge and potential can be controlled by chemical reactions at the surface, for instance in the presence of ionizable groups at the surface, which can release counter-ions into the solution \cite{Markovich2016b}. We will present here the simple and common case of a single-site dissociation process, described by the reaction: 
\begin{equation}
    \ce{AB <--> A^- + B^+}, 
\end{equation}
where $\ce{AB}$ denotes a neutral surface site that can be ionized, $\ce{A^-}$, releasing a counter-ion $\ce{B^+}$ in solution (we arbitrarily consider a negatively charged surface, and monovalent ionized sites). This process is characterized by an equilibrium constant $K_\text{d}$ through:  
\begin{equation}\label{eq:CR_equil}
    K_\text{d} = \frac{[\ce{A^-}][\ce{B^+}]_\text{s}}{[\ce{AB}]}, 
\end{equation}
where $[\ce{A^-}]$, $[\ce{B^+}]_\text{s}$, and $[\ce{AB}]$ denote the concentrations of the species at the surface. 

Denoting $a$ the average distance between the ionizable sites on the surface, one can relate the fraction of ionized groups $\varphi_\text{A}$ to the surface charge: $\varphi_\text{A} = -\Sigma a^2 / e$. 
$\varphi_\text{A}$ is also related to the concentrations of $\ce{AB}$ and $\ce{A^-}$ at the surface through: $[\ce{A^-}]/[\ce{AB}]=\varphi_\text{A}/(1-\varphi_\text{A})$. 

The concentration of $\ce{B^+}$ at the interface can then be expressed as a function of the surface potential, and Eq.~\eqref{eq:CR_equil} provides a relation between the surface charge and the surface potential. Formally, denoting $n_\text{B}^\text{bulk}$ the concentration of $\ce{B^+}$ in bulk, $[\ce{B^+}]_\text{s} = n_\text{B}^\text{bulk} \exp (-\phis)$, and Eq.~\eqref{eq:CR_equil} can be rewritten: 
\begin{equation}
    \varphi_\text{A} = -\frac{\Sigma a^2}{e} = \frac{1}{1+ \frac{1}{K_\text{d}} n_\text{B}^\text{bulk} \exp (-\phis)}. 
\end{equation}

Finally, interfacial electrostatics provides another relation between surface charge and surface potential (e.g. Grahame equation, Eq.~\eqref{eq:grahame_onewall}, for one wall with salt), so that $\Sigma$ and $\phis$ can be uniquely determined by combining the two equations.

\section{Thermodynamics of the EDL}
\label{sec:thermodynamics}

This section describes the thermodynamics of the EDL, i.e., the system comprising the diffuse layer and the charged wall. 
It focuses on the surface excesses \cite{Rowlinson2013} due to the EDL, which will be denoted with an s subscript. 
To simplify the equations, we consider a $Z$:$Z$ electrolyte solution in contact with a single planar wall of area $\mathcal{A}$, where $z \geq 0$ is the distance from the wall, see section~\ref{sec:onewall_salt}. Poisson equation, Eq.~\eqref{eq:Poisson_onewall}, and the electrostatic boundary condition following from global electroneutrality, Eq.~\eqref{eq:interface_ES_onewall}, are obeyed. 

In this section only, surface tension will be noted simply as $\gamma$, not to be confused with the $\gamma$ parameter introduced in previous sections. In all other sections, surface tension is referred to as $\gamma_\text{EDL}$.

\subsection{Grand potential excess of the EDL}

The electrostatic energy per unit area of the EDL is \cite{Overbeek1990,Gupta2020}: 
\begin{equation}
    \begin{split}
        \frac{U_\text{el}}{\mathcal{A}} &= \frac{1}{2} \int_0^\infty \rhoe(z) V(z) \dd z + \frac{1}{2} \Sigma \Vs \\
        &= \frac{\eps}{2} \int_0^\infty \left( \frac{\dd V}{\dd z} \right)^2 \dd z = \mathcal{E}.  
    \end{split}
\end{equation}
To obtain the total surface energy excess due to the EDL, one must subtract the energy required to charge the wall, $U_\text{charge}$ \cite{Overbeek1990,Gupta2020}. When the wall is maintained at a fixed surface potential, $U_\text{charge}/\mathcal{A} = \Sigma \Vs$, while when working at fixed surface charge, one can set $U_\text{charge} = 0$ (we will come back to this point later).   

One can then write the surface entropy excess, in the dilute solution limit \cite{Markovich2016a,Gupta2020}: 
\begin{multline}\label{eq:S_s}
    \frac{S_\text{s}}{\mathcal{A}} = - k_\text{B} \int_0^\infty \left[ n_+(z) \ln (a^3 n_+(z))  - n_+(z) \right. \\
    + n_-(z) \ln (a^3 n_-(z)) - n_-(z) \\
    \left. - 2 \ns \ln (a^3 \ns) + 2 \ns \right] \dd z,   
\end{multline}
where $a$ is a microscopic length scale, defining a reference close-packing density, $a^{-3}$. 
Also in the dilute solution limit, the electrochemical potential of the ions is:
\begin{equation}
    \mu_\pm = \pm q V(z) + k_\text{B}T \ln \left( a^3 n_\pm(z) \right) = k_\text{B}T \ln \left( a^3 \ns \right), 
\end{equation}
so that $\mu_+ = \mu_- = \mu_0$. 

Finally, one can compute the surface Grand potential excess due to the EDL: 
\begin{equation}\label{eq:Omega_s_tmp}
    \Omega_\text{s} = U_\text{el} - U_\text{charge} - T S_\text{s} - \mu_0 \Gamma \mathcal{A},  
\end{equation}
where $\Gamma = \int_0^\infty ( n_+ + n_- - 2 \ns ) \dd z$ is the surface adsorption of ions (per unit area). 

Equation~\eqref{eq:Omega_s_tmp} can be rewritten: 
\begin{multline}\label{eq:Omega_s}
    \Omega_\text{s} = U_\text{el} - U_\text{charge} + 
    \mathcal{A} k_\text{B}T \int_0^\infty \left[ n_+(z) \ln \left( \frac{n_+(z)}{n_0} \right) \right. \\
    \left. - n_+(z) + n_-(z) \ln \left( \frac{n_-(z)}{n_0} \right) - n_-(z) + 2\ns \right] \dd z.  
\end{multline}
This expression can also be obtained, and extended, using classical density functional theory \cite{Hansen1990,Hartel2017,Cats2022}. 

\subsection{Constant surface potential}

One can evaluate the expression of $\Omega_\text{s}$, Eq.~\eqref{eq:Omega_s}, within the PB framework. 
Working at constant surface potential, $U_\text{charge}/\mathcal{A} = \Sigma \Vs$, so that: 
\begin{multline*}
    \Omega_\text{s} = \mathcal{E} \mathcal{A} - \Vs \Sigma \mathcal{A} \\
     + \mathcal{A} k_\text{B}T \int_0^\infty \left[ -q (n_+ - n_-) V - (n_+ + n_- - 2\ns) \right] \dd z,  
\end{multline*}
which simplifies into: 
\begin{equation}
    \Omega_\text{s} = -2 \mathcal{E} \mathcal{A}.
\end{equation}

The surface Grand potential excess of the EDL, $\Omega_\text{s}$, depends on $T$, $\mathcal{A}$, $\mu_0$, and $\Vs$. Because $\Omega_\text{s}$ is surface-extensive, one can write \cite{Cats2021}: 
\begin{equation}
    \Omega_\text{s} = \gamma(T,\mu_0,\Vs) \mathcal{A}, 
\end{equation}
with the surface tension $\gamma$ given by: 
\begin{equation}
        \gamma = \left( \frac{\partial \Omega_\text{s}}{\partial \mathcal{A}} \right)_{T,\mu_0,\Vs} = -2 \mathcal{E}. 
\end{equation}
The differential of $\Omega_\text{s}$ is then: 
\begin{equation}\label{eq:diff_Omega_s}
    \begin{split}
        \dd \Omega_\text{s} &= \gamma \dd \mathcal{A} + \mathcal{A} \dd \gamma \\
        &= -\mathcal{A} s_\text{s} \dd T + \gamma \dd \mathcal{A} - \mathcal{A} \Gamma \dd \mu_0 - \mathcal{A} \Sigma \dd \Vs,
    \end{split}
\end{equation} 
denoting $s_\text{s}$ the surface density of entropy excess. 
From Eq.~\eqref{eq:diff_Omega_s}, one can write a ``surface Gibbs-Duhem'' equation, sometimes referred to as Lippmann equation: 
\begin{equation}
        \dd \gamma = -s_\text{s} \dd T - \Gamma \dd \mu_0 - \Sigma \dd \Vs. 
\end{equation}
One can then compute the surface excesses: 
\begin{equation}\label{eq:s_s_thermo}
\begin{split}
        s_\text{s} &= -\left( \frac{\partial \gamma}{\partial T} \right)_{\mu_0,\Vs} \\
        &= \frac{1}{T} \left\{ 3 \mathcal{E} - \Sigma \Vs \right\} - \frac{1}{T} \mu_0 \Gamma + \mathcal{E} \frac{1}{\eps} \frac{\dd \eps}{\dd T}
\end{split}
\end{equation}
\begin{equation}\label{eq:Gamma_thermo}
        \Gamma = -\left( \frac{\partial \gamma}{\partial \mu_0} \right)_{T,\Vs} = -\beta \ns \left( \frac{\partial \gamma}{\partial \ns} \right)_{T,\Vs}= \beta \mathcal{E} , 
\end{equation}
and verify the validity of: 
\begin{equation}
        \Sigma = -\left( \frac{\partial \gamma}{\partial \Vs} \right)_{T,\mu_0}.  
\end{equation}

As a first side note, in the expression of $s_\text{s}$, the first two terms correspond to the mixing entropy excess, Eq.~\eqref{eq:S_s}, but the third term arises from $U_\text{el}$. Indeed, the electrostatic energy is in fact a free energy, containing both an energy and an entropy contribution of the solvent \cite{Overbeek1990}. 

As a second side note, Eq.~\eqref{eq:s_s_thermo} can be used to compute the Marangoni surface stress, i.e. the surface tension gradient, due to the EDL under a thermal gradient parallel to the interface. Working for instance with an imposed bulk salt concentration $\ns$:  
\begin{equation}\label{eq:surf_tension_gradient_temp}
\begin{split}
    \nabla \gamma &= \left( \frac{\partial \gamma}{\partial T} \right)_{\ns,\Vs} \times \nabla T
    = - \left( s_\text{s} + \frac{1}{T} \mu_0 \Gamma \right) \nabla T \\
    &= -(3 \mathcal{E} - \Sigma \Vs) \frac{\nabla T}{T} - \mathcal{E} \frac{\nabla \eps}{\eps}. 
\end{split}
\end{equation}
Similarly Eq.~\eqref{eq:Gamma_thermo} can be used in the case of an isothermal salt concentration gradient: 
\begin{equation}\label{eq:surf_tension_gradient_salt}
    \nabla \gamma = -\Gamma \nabla \mu_0 = - \Gamma \frac{1}{\beta} \frac{\nabla \ns}{\ns} . 
\end{equation}
These expressions identify with the integral over the EDL of the local pressure gradient within the  standard PB description of osmotic flows, see section~\ref{sec:EK_effects}.

\subsection{Constant surface charge density}

To describe a wall with a fixed surface charge density, one can redefine the Grand potential through a Legendre transform: 
\begin{equation}
    \Omega'_\text{s} = \Omega_\text{s} + \Vs \Sigma \mathcal{A} = (-2 \mathcal{E} + \Vs \Sigma) \mathcal{A}.
\end{equation}
Note that this is equivalent to setting $U_\text{charge}=0$ in the general expression of $\Omega_\text{s}$, Eq.~\eqref{eq:Omega_s}; see above discussion on the energy excess of the EDL. 

Because the new natural variable is $\Sigma$ and not the total surface charge $Q_\text{s} = \Sigma \mathcal{A}$, one gets: 
\begin{equation}
    \left( \frac{\partial \Omega'_\text{s}}{\partial \mathcal{A}} \right)_{T,\mu_0,\Sigma} = \gamma + \Vs \Sigma.
\end{equation}
Denoting $\gamma' = \gamma + \Vs \Sigma$, 
one can then use again the surface-extensivity of $\Omega'_\text{s}$ to derive the surface Gibbs-Duhem equation: 
\begin{equation}
        \dd \gamma' = -s_\text{s} \dd T - \Gamma \dd \mu_0 - \Vs \dd \Sigma,  
\end{equation}
from which one can compute the surface excesses (which do not depend on the boundary condition): 
\begin{align}
        s_\text{s} &= -\left( \frac{\partial \gamma'}{\partial T} \right)_{\mu_0,\Sigma} = -\left( \frac{\partial \gamma}{\partial T} \right)_{\mu_0,\Vs} \\
        \Gamma &= -\left( \frac{\partial \gamma'}{\partial \mu_0} \right)_{T,\Sigma} = -\left( \frac{\partial \gamma}{\partial \mu_0} \right)_{T,\Vs}, 
\end{align}
and verify the validity of: 
\begin{equation}
        \Vs = -\left( \frac{\partial \gamma'}{\partial \Sigma} \right)_{T,\mu_0}.  
\end{equation}

\section{Application of the PB framework: electrokinetic effects}
\label{sec:EK_effects}

Electrokinetic (EK) effects in a broad sense refer to the coupling between different types of transport (hydrodynamic, electrical, solutal, thermal) in micro and nanofluidic systems. The EDL plays a central role in EK effects, and a description of the EDL at the PB level can provide excellent estimates of the EK response, and serve as a basis for more refined models. Hence, the PB equation is still at the core of the modern description of EK effects \cite{Mouterde2018,Vinogradova2020,Vinogradova2022,Emmerich2022,Mangaud2021,Allemand2023,Pandey2023,Vinogradova2023}.

Detailed articles and reviews exist on the subject \cite{Anderson1989,Gravelle2016review,Marbach2019,Herrero2022a}. Here we will simply illustrate how the EK response can be derived within the PB framework, focusing on the so-called osmotic flows and on the electric conductivity.

\subsection{Osmotic flows}

Osmotic flows are generated by non-hydrodynamic forcing along interfaces (e.g., electric field for electro-osmosis, solute concentration gradient for diffusio-osmosis, thermal gradient for thermo-osmosis). 
Indeed, as detailed below, in the EDL where the liquid interacts with the wall, external fields produce local pressure gradients, which drives the liquid into motion.
The standard description of osmotic flows is based on Stokes equation, i.e., on continuum hydrodynamics at low Reynolds numbers, considering a newtonian incompressible liquid with a homogeneous viscosity $\eta$, driven by a local pressure gradient / force density in the vicinity of the wall \cite{Anderson1989,Gravelle2016review,Marbach2019,Herrero2022a}. 

Let's consider a $Z$:$Z$ electrolyte solution in contact with a single planar wall as in Sec.~\ref{sec:onewall_salt}, denoting $z$ the distance to the wall. When a force density $f(z)$ (vanishing in the bulk liquid far from the wall) is applied to the liquid along the $x$ direction parallel to the wall, the Stokes equation writes: 
\begin{equation}
\label{eq:stokes_scalar}
- \eta \frac{\dd^2 v}{\dd z^2} = f(z), 
\end{equation}
with $v(z)$ the osmotic velocity profile. Stokes equation can be solved once the hydrodynamic boundary condition is known. At the macro and micro scales, the standard no-slip boundary condition can be used. At the nanoscale however, depending on the amplitude of interfacial friction, there can be a stagnant layer of liquid of thickness $z_\text{s}$ (being on the order of one molecular diameter), or the liquid can slip on the wall \cite{Herrero2022a,Hadjiconstantinou2024}. The latter case is described by the partial slip boundary condition: 
\begin{equation}
    v(0) = b \left. \frac{\dd v}{\dd z} \right|_{z=0} , 
\end{equation}
quantified by the so-called slip length $b$. In the following, we will focus on the case of a slipping wall. The formulas for a no-slip wall are simply obtained by setting $b=0$.

One can solve Eq.~\eqref{eq:stokes_scalar} with the slip boundary condition and a vanishing shear rate far from the wall to obtain the osmotic velocity profile: 
\begin{equation}
    v(z) = \frac{1}{\eta} \qty[ \int_0^z \dd z' \int_{z'}^{\infty} f(z'') \dd z'' + b \int_0^\infty f(z) \dd z ].
    \label{eq:vx_z}
\end{equation}
The osmotic velocity $v_{\rm osm}^\infty$ far from the wall is then:
\begin{equation}
    v^\infty_{\rm osm} = \lim_{z \to \infty} v(z) = \frac{1}{\eta} \int_0^\infty (z+b) f(z) \dd z.
    \label{eq:vosm}
\end{equation}
We will now apply these general formulas to the different types of osmotic flows.

\subsubsection{Electro-osmosis and zeta potential}

When an external electric field $E$ is applied parallel to the interface, the force density is $f(z) = \rhoe(z) E$. Accordingly, the electro-osmotic velocity profile is: 
\begin{equation}
    v_\text{eo}(z) = \frac{1}{\eta} \qty[ \int_0^z \dd z' \int_{z'}^{\infty} \rhoe(z) \dd z'' - b \Sigma ] \times E,
    \label{eq:veo_z}
\end{equation}
and the electro-osmotic velocity far from the wall is: 
\begin{equation}
    v^\infty_{\rm eo} = \frac{1}{\eta} \int_0^\infty (z+b) \rhoe(z) \dd z \times E.
    \label{eq:veo}
\end{equation}
Using Poisson equation, Eq.~\eqref{eq:Poisson_onewall}, and the electrostatic boundary condition, Eq.~\eqref{eq:interface_ES_onewall}, these expressions can be simplified: 
\begin{multline}
v_{\rm eo}(z)
= - \frac{\eps}{\eta} \qty(\Vs - V(z) + \frac{\Sigma b}{\eps}) \times E \\
    = - \frac{\eps \Vs}{\eta} \qty(1 - \frac{V(z)}{\Vs} + \frac{b}{\lambda}) \times E,
\end{multline} 
and
\begin{equation}
    v^\infty_{\rm eo} 
    = - \frac{\eps}{\eta} \qty(\Vs + \frac{\Sigma b}{\eps}) \times E
    = - \frac{\eps \Vs}{\eta} \qty(1 + \frac{b}{\lambda}) \times E, 
\end{equation}
where $\lambda = -\Vs/(\mathrm{d}V/\mathrm{d}z|_{z=0})$ is the effective thickness of the EDL defined in Sec.~\ref{sec:onewall_salt_integrals}. 

The electro-osmotic response is commonly quantified in terms of an effective potential $\zeta$, defined from the Helmholtz–Smoluchowski equation \cite{Delgado2007,Hartkamp2018}: $v_{\rm eo}^\infty = - \frac{\eps \zeta}{\eta}E$, so that: 
\begin{equation*}
    \zeta
    = \Vs + \frac{\Sigma b}{\eps}
    = \Vs \qty( 1 + \frac{b}{\lambda_{\rm eff}} ).
\end{equation*}
This equation emphasizes that the zeta potential obtained through macroscopic EK measurements depends on both interfacial electrostatics and hydrodynamics, and does not necessarily identify with the surface potential $\Vs$ \cite{Joly2004,Lyklema2011,Maduar2015,Hartkamp2018}.

\subsubsection{Diffusio-osmosis}

When a salt concentration gradient $\nabla \ns$ is applied in the bulk liquid along the $x$ direction parallel to the interface, one can compute the corresponding force density driving the flow by assuming local thermal equilibrium \cite{Ramirez-Hinestrosa2021,Anderson1989,Herrero2022a}. 
Indeed, the Gibbs-Duhem relation then provides a link between the thermodynamic gradients along the $x$ direction at different distances $z$ from the wall. Ignoring the possible layering of the solvent close to the wall, the force density -- which identifies with the local pressure gradient -- then writes: 
\begin{equation*}
    f(z) = \qty[ n_+(z) + n_-(z) - 2n_0 ] \times \left(-\frac{1}{\beta} \frac{\nabla \ns}{\ns} \right)
\end{equation*}
Accordingly, the diffusio-osmotic velocity far from the interface is: 
\begin{multline}
    v^\infty_{\rm do} = \frac{1}{\eta} \int_0^\infty (z+b) \qty[ n_+(z) + n_-(z) - 2n_0 ] \dd z \\
    \times \left(-\frac{1}{\beta} \frac{\nabla \ns}{\ns} \right).
    \label{eq:vdo}
\end{multline}
In the PB framework, this expression simplifies into: 
\begin{equation}\label{eq:vdobulk_PB_slip}
    v^\infty_{\rm do} = \frac{1}{\eta} \left\{ - \frac{1}{2\pi\lB} \ln \left( 1 - \gamma^2\right) + \Gamma b \right\}  \times \left(-\frac{1}{\beta} \frac{\nabla \ns}{\ns} \right), 
\end{equation}
with $\Gamma$ the adsorption defined in Sec.~\ref{sec:onewall_salt_integrals}.
In this expression, the slip contribution can be rewritten in terms of the surface tension gradient induced by an isothermal salt concentration gradient, see Eq.~\eqref{eq:surf_tension_gradient_salt} in Sec.~\ref{sec:thermodynamics}:
\begin{equation}
    v^{\infty}_{\rm do,slip} = \frac{b}{\eta} \left\{ - \Gamma \frac{1}{\beta} \frac{\nabla \ns}{\ns} \right\} = \frac{b}{\eta}  \nabla \gamma_\text{EDL}. 
\end{equation}
Indeed, the slip contribution can be obtained by balancing the surface tension gradient, $\nabla \gamma_\text{EDL}$, with the interfacial friction stress, $(\eta/b) v^{\infty}_{\rm do,slip}$.

Note that an additional flow can be generated under a salt concentration gradient, when cations and anions have a different diffusion coefficient, $D_+ \neq D_-$, and when the channel boundary conditions impose that there is no electrical current along the flow direction in the bulk liquid \cite{Anderson1989,Herrero2022a,Lee2014b}. 
Indeed, in that case a so-called diffusion electric field $E_0$ appears to avoid charge separation, which one can compute by writing that the bulk electric current vanishes: 
\begin{equation}
    E_0 = \delta \frac{1}{\beta q} \frac{\nabla \ns}{\ns},     
\end{equation}
with $\delta = (D_+ - D_-)/(D_+ + D_-)$. 
The diffusion electric field creates an electro-osmotic flow (see previous section), which adds to the diffusio-osmotic flow.

\subsubsection{Thermo-osmosis}

Following the same approach as for diffusio-osmosis, one can compute the force density under a temperature gradient $\nabla T$ \cite{Wurger2010,Ganti2017,Ouadfel2024}: 

\begin{multline}
    f(z)=-\left\{ \rhoe(z) V(z) + \frac{1}{\beta} \qty[ n_+(z) + n_-(z) - 2n_0 ] \right\} \frac{\nabla T}{T} \\
    - \frac{E(z)^2}{2}\nabla\varepsilon,
\end{multline}
where the second term arises from the gradient of dielectric permittivity $\varepsilon$ induced by temperature. Here, $E(z)$ is the electric field, which is approximately equal to its transverse component $E_z(z)$ close to an electrically charged surface. 

In the PB framework, the force density becomes:
\begin{multline}
    f(z)=-\left[-\varepsilon V(z)\dv[2]{V}{z}+\frac{\varepsilon}{2}\left(\dv{V}{z}\right)^2\right]\frac{\nabla T}{T}\\
    -\frac{1}{2}\left(\dv{V}{z}\right)^2\nabla \varepsilon,  
\end{multline} 
and the thermo-osmotic velocity far from the interface writes: 
\begin{equation}
    v_\mathrm{to}^{\infty}=v_{\mathrm{to},\nabla T}^{\infty}+v_{\mathrm{to},\nabla\varepsilon}^{\infty},
\end{equation}
with
\begin{multline}
    v_{\mathrm{to},\nabla T}^{\infty}=\frac{-\nabla T/T}{\eta} \biggl\{ \frac{1}{2\pi\ell_{\mathrm{B}}\beta} \left[ -3 \ln \left( 1 - \gamma^2\right) - \frac{\phis^2}{4} \right] \\ 
    + b \left[ 3 \mathcal{E} - \Sigma \Vs \right] \biggr\} 
\end{multline}
\begin{equation}\label{eq:vto_eps}
    v_{\mathrm{to},\nabla\varepsilon}^{\infty} = \frac{-\nabla\varepsilon/\varepsilon}{\eta} \left\{ - \frac{1}{2\pi\lB\beta} \ln \left( 1 - \gamma^2\right) + \mathcal{E} b \right\}, 
\end{equation}
with $\mathcal{E}$ the electrostatic energy defined in Sec.~\ref{sec:onewall_salt_integrals}.
As for diffusio-osmosis, the slip contribution to the thermo-osmotic velocity can be rewritten in terms of the surface tension gradient induced by a thermal gradient, see Eq.~\eqref{eq:surf_tension_gradient_temp} in Sec.~\ref{sec:thermodynamics}:
\begin{equation}
    v^{\infty}_{\rm to,slip} = \frac{b}{\eta} \left\{ -(3 \mathcal{E} - \Sigma \Vs) \frac{\nabla T}{T} - \mathcal{E} \frac{\nabla \eps}{\eps} \right\} = \frac{b}{\eta}  \nabla \gamma_\text{EDL}. 
\end{equation}

Note that the permittivity gradient can be expressed as a function of the thermal gradient \cite{Wurger2010}:
\begin{equation}
    \frac{\nabla\varepsilon}{\varepsilon}=-\tau\frac{\nabla T}{T},
\end{equation}
with $\tau=1.4$ for water at room temperature.

\subsection{Electric conductivity}

Using the same description of hydrodynamics as for osmotic flows, and assuming a fixed homogeneous mobility for the ions, the PB equation can provide a simple expression for the average conductivity of micro and nanochannels \cite{Herrero2022a}. 

Let's consider, for instance, the slit channel represented in Fig.~\ref{fig:PB_figs}(c), in the thin EDL limit, $\debye \ll d$. An electric field $E$ applied along the channel will induce an electric current. The electric current density writes: 
\begin{multline}
  j_e(z) = q \left[ n_+(z) v_+(z) - n_-(z) v_-(z) \right], \\ 
  \text{with}\quad v_\pm(z) = \pm q \mu_\pm E + v_\text{eo} (z),
\end{multline}
with $\mu_\pm$ the ion mobility and $v_\text{eo}(z)$ the electro-osmotic velocity profile.
The electric conductivity $\sigma$ (averaged over the channel section) then writes (detailed calculations can be found in Refs.~\citenum{Bocquet2010,Balme2015,Werkhoven2020}): 
\begin{multline}
  \sigma = 
  \frac{2}{E d} \int_0^{d/2} j_e(z) \,\dd z 
  = 2 q^2 \bar{\mu} \ns \\
  +\frac{1}{d} \left\{ 2 q^2 \bar{\mu} \beta\mathcal{E} - 2q \Delta\mu \Sigma + \frac{q^2 \beta\mathcal{E}}{\pi \lB \eta} + \frac{2 b \Sigma^2}{\eta} \right\}, \label{eq:conductivity}
\end{multline}
where $\bar{\mu} = \frac{\mu_+ + \mu_-}{2}$, $\Delta \mu = \frac{\mu_+ - \mu_-}{2}$, and $\mathcal{E}$ is the electrostatic energy (per unit area) of the EDL on one wall, see section~\ref{sec:onewall_salt_integrals}.
The first term in Eq.~\eqref{eq:conductivity} is the bulk ionic conductivity, and the terms between the curly brackets are surface contributions, respectively, from ionic mobility (with a first term controlled by the average mobility $\bar{\mu}$ and a second controlled by the mobility asymmetry $\Delta \mu$), and from the advection by the electro-osmotic flow (with a first term related to the no-slip part of the flow, and an additional term due to slip).

\section{Limits of the PB framework}
\label{sec:limits}

In the first molecular layers of liquid close to the wall, the hypotheses underlying the PB equation generally fail. The reader can find detailed discussions on the limits of the PB theory in, e.g., Refs.~\citenum{LyklemaBook,Hunter2001,IsraelachviliBook,Andelman1995,Markovich2016a,Delgado2007,Schoch2008,Bonthuis2013,Hartkamp2018,Faucher2019,Kavokine2021,Becker2024}. 

Here we will simply present a few criteria, starting with one for the validity of the mean field approximation. 
The importance of ionic correlations can be quantified by the plasma parameter $\Gamma$, see Refs.~\citenum{Levin2002,Levin2003,Joly2006}, which compares the typical interaction energy between two ions and the thermal energy $k_\mathrm{B}T$: 
\begin{equation}
\Gamma = \frac{\beta q^2}{4\pi \eps d_\mathrm{ion}} = \frac{\lB}{d_\mathrm{ion}},
\end{equation}
where $d_\mathrm{ion}$ is the typical inter-ionic distance. 

This typical inter-ionic distance can be related to surface or bulk properties, imposing respectively a critical surface charge density $|\Sigma|^\mathrm{c}$ or a critical salt concentration $n_0^\mathrm{c}$ above which the applicability of the PB framework should be taken with care.

At the surface, assuming that counter-ions organize into a monolayer screening the surface charge, $1/d_\mathrm{ion}^{\,2}=|\Sigma|/q$, and 
$\Gamma^\text{surface} = \sqrt{|\Sigma| \lB^2 / q}$. 
Ionic correlations cannot be neglected when $\Gamma^\text{surface} > 1$, see Refs.~\citenum{Levin2002,Levin2003}, corresponding to: 
\begin{equation}
    |\Sigma| > |\Sigma|^\mathrm{c} = \frac{q}{\lB^2} = \frac{(4\pi\eps/\beta)^2}{(Ze)^3}, 
\end{equation}
where the last expression highlights the strong impact of the ion valence $Z$ on $\abs{\Sigma^\mathrm{c}}$.
For monovalent ions in water at $300\,$K, $\abs{\Sigma^\mathrm{c}} \sim 330\,$mC/m$^2$, but for divalent ions it drops to $\sim 40\,$mC/m$^2$. 
As a side note, $\Gamma^\text{surface}$ is closely related to the coupling parameter $\Xi$ introduced in the strong coupling theory to quantify the importance of ionic correlations \cite{Netz2001}: $\Gamma^\text{surface} = \sqrt{\Xi/(2\pi)}$.

In bulk, because the total ionic concentration is twice the salt concentration, the inter-ionic distance is $d_\mathrm{ion}=(2 n_0)^{-1/3}$, and $\Gamma^\text{bulk} = (2 n_0 \lB^3)^{1/3}$. Ionic correlations cannot be neglected when $\Gamma^\text{bulk} > 1$, see Refs.~\citenum{Levin2002,Levin2003}, corresponding to: 
\begin{equation}
    n_0 > n_0^\mathrm{c}=\frac{1}{2\lB^3} = \frac{(4\pi\eps/\beta)^3}{2(Ze)^6}.
\end{equation}
For a monovalent salt in water at $300\,$K, $n_0^\mathrm{c}\sim 2\,$M, but for a divalent salt it drops to $\sim 30$\,mM. 

As a side note, for monovalent ions in water at room temperature, $\lB \sim 7\,$\AA{} is greater than the ionic size, so that there will be no steric repulsion effects when $\Gamma = \lB/d_\mathrm{ions} < 1$. 

The size of the system can also play a role in the suitability of a theoretical description within the PB framework. For instance, for extremely small pores (of radius on the order of nm), when the surrounding medium has a much lower permittivity than the channel, the potential at intermediate distances $x$ along the channel is similar to the 1D Coulomb potential, due to dielectric confinement \cite{Kavokine2021}:
\begin{equation}
    V(x) \approx \frac{q\alpha}{2\pi \eps R}e^{-x/\alpha R},
\end{equation}
where $\alpha$ is a numerical coefficient dependent on the ratio of $\epsilon$ with the dielectric permittivity of the medium surrounding the channel (e.g., $\alpha = 6.3$ for a ratio of 40). 
For extreme confinement, one should also consider that the inter-ionic distance is given by the relation $d_{\rm ions}=(\pi R^2 \rho)^{-1}$, with $\rho$ the ion concentration inside the channel. 
Then, because $\Gamma = \beta q V(d_{\rm ions})$, and defining the critical ion concentration $\rho^{\rm c}$ when $\Gamma = 1$, one obtains:
\begin{equation}
    \rho^{\rm c} = \frac{1}{\pi R^3 \alpha \ln(\frac{2\lB \alpha}{R})}.
\end{equation}
Considering monovalent ions in water at 300\,K (so $\lB\sim 7$\,\AA), using $\alpha \sim 6.3$ and a infinitely long pore of radius $R=2$\,nm, then $\rho^{\rm c} \sim 7$\,mM.

Aside of the channel size, the channel length $L$ may also play a role in determining the limits of validity of the PB framework, due to the possible breakdown of electroneutrality in the pore \cite{levy2020}. For instance, considering a cylindrical channel of radius $R$ and length $L$, at low surface charges, electroneutrality will breakdown for 
\[
\debye > \debye^{\rm c} = R \sqrt{ 2 \ln(\frac{L}{2R}) },
\]
or in terms of salt concentration:
\begin{equation}
    n_0 < n_0^{\rm c} = \frac{1}{16 \pi \lB R^2 \ln(\frac{L}{2R})}.
\end{equation}
The critical salt concentration increases with decreasing $L/R$ ratio and with decreasing $R$. 
For a monovalent salt in water at 300\,K, with e.g. $R=2$\,nm and $L=10$\,nm, $n_0^{\rm c} \sim 13$\,mM.

Finally, note that the criteria presented above are based on electrostatic considerations only. Other criteria exist, for instance considering the association of ions into Bjerrum pairs, which also invalidates the mean-field approach \cite{Valeriani2010}. The quasi-chemical ion-pairing theory proposed by Bjerrum can be used to derive a critical salt density over which ion pairing occurs significantly. The theory involves an equilibrium constant $K$, for which an approximate expression can be obtained as a function of $\lB/\sigma$, where $\sigma$ is the ionic diameter, in the limit of $\lB/\sigma \gg 1$: 
\begin{equation}\label{eq:Bjerrum_pairs}
    K \approx 4\pi \sigma^3 e^{\lB/\sigma} \left[ \frac{\sigma}{\lB} + 2 \left(\frac{\sigma}{\lB}\right)^2 + 2 \left(\frac{\sigma}{\lB}\right)^3 \right]. 
\end{equation}
Ion pairing cannot be neglected when $\ns > 4/K$. Interestingly, in the range of validity of Eq.~\eqref{eq:Bjerrum_pairs}, this criterion provides similar orders of magnitude for the critical ion density as the one presented above, except at very large Bjerrum lengths ($\lB/\sigma>7$) where this new criterion becomes more stringent.

\begin{acknowledgments}
The authors thank E. Trizac, O. Vinogradova and Ren\'e van Roij for their stimulating feedback.
\end{acknowledgments}

\section*{Author Declarations}

The authors have no conflicts to disclose. 

\section*{Data availability}

Data sharing is not applicable to this article as no new data were created or analyzed in this study.

%

\appendix
\onecolumngrid

\section{Numerical solution with Python}

The Python codes listed below solve numerically the PB equation in the slit and cylindrical geometries, using the \verb!solve_bvp! function of the \verb!SciPy! package. The script files are provided as supplementary material. Both scripts can be downloaded in the GitHub repository: \url{https://github.com/ceciherr/PBformulary} .

\subsection{Slit channel} 
\label{sec:pb_slit_num}

This code computes and plots the auxiliary reduced potential $\psi= -\text{sgn}(\Sigma) \times \phi$ as a function of the reduced position $z/d$ in a slit channel with walls located at $z=-d/2$ and $z=d/2$ (see Sec.~\ref{sec:twowalls_salt}), for a broad range of values of $r_1 = d/\debye$ and $r_2 = \debye/\lGC$. 

To use the \verb!solve_bvp! function, one needs to convert the PB equation into a first order system of ordinary differential equations subject to two-point boundary conditions: 
\begin{equation*}
    \frac{\dd^2 \psi}{\dd z^2} = \frac{1}{\debye^2} \sinh\{\psi\} 
    \quad \text{with} \quad 
    \left.\frac{\dd \psi}{\dd z}\right|_{z=0} = 0
    \quad \text{and} \quad 
    \left.\frac{\dd \psi}{\dd z}\right|_{z=d/2} = -\frac{2}{\lGC}
\end{equation*}
\begin{equation*}
    x = z/d, \quad y_0 = \psi, \quad r_1 = d/\debye, \quad r_2 = \debye/\lGC \quad \Rightarrow
    \left\{
    \begin{aligned}
    &y_0' = y_1\\
    &y_1' = r_1^2 \sinh(y_0)\end{aligned}\right.
    \quad \text{with} \quad
    \left\{\begin{aligned}
    &y_1(0) = 0\\
    &y_1(1/2) = -2r_1 r_2\end{aligned}\right.
\end{equation*}

\subsubsection*{Listing}

\begin{verbatim}
#!/usr/bin/env python3
# -*- coding: utf-8 -*-

r1 = 5.
r2 = 5.

import numpy as np
import matplotlib.pyplot as plt
from scipy.integrate import solve_bvp

# y0 = psi
# x = z/d

# Right-hand side of the system: 
# y0' = y1
# y1' = r1^2 sinh(y0)    
def fun(x, y):
    return np.vstack((y[1], r1**2*np.sinh(y[0])))

# Boundary conditions:
# y1(left) = 0
# y1(right)+2r1r2 = 0
def bc(ya, yb):
    return np.array([ya[1], yb[1]+2.0*r1*r2])

# Solving for x in [0;0.5]
x = np.linspace(0., 0.5, 10000)
y = np.zeros((2, x.size))

# Initial guess for the potential profile
r1init = 1.
r3init = 3.135322030076839  # r2init = 1000
y[0] = 2.*np.log(r1init/(2.*r3init)*np.cos(r3init*x))    

# Calling solve_bvp
res = solve_bvp(fun, bc, x, y, tol=1.e-6, max_nodes=5000000)

x_plot = np.linspace(0., 0.5, 500)
y_plot = res.sol(x_plot)[0]
plt.plot(x_plot, y_plot, label='numerical')

plt.legend()
plt.xlabel(r"$z/d$")
plt.ylabel(r"$\psi$")

plt.show()
\end{verbatim}


\subsection{Cylindrical channel} 
\label{sec:pb_cyl_num}

This code computes and plots the auxiliary reduced potential $\psi= -\text{sgn}(\Sigma) \times \phi$ as a function of the reduced position $r/R$ in a cylindrical channel of radius $R$ (see Sec.~\ref{sec:cyl_salt}), for a broad range of values of $r_1 = R/\debye$ and $r_2 = \debye/\lGC$. 

To use the \verb!solve_bvp! function, one needs to convert the PB equation into a first order system of ordinary differential equations subject to two-point boundary conditions: 
\begin{equation*}
    \frac{\dd^2 \psi}{\dd r^2} + \frac{1}{r} \frac{\dd \psi}{\dd r} = \frac{1}{\debye^2} \sinh\{\psi\} 
    \quad \text{with} \quad 
    \left.\frac{\dd \psi}{\dd r}\right|_{r=0} = 0
    \quad \text{and} \quad 
    \left.\frac{\dd \psi}{\dd r}\right|_{r=R} = -\frac{2}{\lGC}
\end{equation*}
\begin{equation*}
    x = r/R, \quad y_0 = \psi, \quad r_1 = R/\debye, \quad r_2 = \debye/\lGC \quad \Rightarrow
    \left\{
    \begin{aligned}
    &y_0' = y_1\\
    &y_1' = r_1^2 \sinh(y_0) - y_1/x\end{aligned}\right.
    \quad \text{with} \quad
    \left\{\begin{aligned}
    &y_1(0) = 0\\
    &y_1(1) = -2r_1 r_2\end{aligned}\right.
\end{equation*}

\subsubsection*{Listing}

\begin{verbatim}
#!/usr/bin/env python3
# -*- coding: utf-8 -*-

r1 = 5.
r2 = 5.

import numpy as np
import matplotlib.pyplot as plt
from scipy.integrate import solve_bvp

# y0 = psi
# x = r/R

# Right-hand side of the system: 
# y0' = y1
# y1' = r1^2 sinh(y0)-y1/x
def fun(x, y):
    return np.vstack((y[1], r1**2*np.sinh(y[0])-y[1]/x))

# Boundary conditions:
# y1(left) = 0
# y1(right)+2r1r2 = 0
def bc(ya, yb):
    return np.array([ya[1], yb[1]+2.0*r1*r2])

# Solving for x in ]0;1]
x = np.linspace(1.e-6, 1., 10000)
y = np.zeros((2, x.size))

# Initial guess for the potential profile
r1init = 1.
r3init = 1.9980029950087341  # r2init = 1000
y[0] = 2.*np.log(r1init/(2.*r3init)*(1.-(r3init*x/2)**2))  

# Calling solve_bvp
res = solve_bvp(fun, bc, x, y, tol=1.e-6, max_nodes=5000000)

x_plot = np.linspace(0., 1., 500)
y_plot = res.sol(x_plot)[0]
plt.plot(x_plot, y_plot, label='numerical')

plt.legend()
plt.xlabel(r"$r/R$")
plt.ylabel(r"$\psi$")

plt.show()
\end{verbatim}


\end{document}